\PassOptionsToPackage{unicode}{hyperref}
\PassOptionsToPackage{hyphens}{url}
\PassOptionsToPackage{dvipsnames,svgnames,x11names}{xcolor}
\documentclass[
  11pt,
]{article}
\usepackage{amsmath,amssymb}
\usepackage{lmodern}
\usepackage{iftex}
\ifPDFTeX
  \usepackage[T1]{fontenc}
  \usepackage[utf8]{inputenc}
  \usepackage{textcomp} 
\else 
  \usepackage{unicode-math}
  \defaultfontfeatures{Scale=MatchLowercase}
  \defaultfontfeatures[\rmfamily]{Ligatures=TeX,Scale=1}
\fi
\IfFileExists{upquote.sty}{\usepackage{upquote}}{}
\IfFileExists{microtype.sty}{
  \usepackage[]{microtype}
  \UseMicrotypeSet[protrusion]{basicmath} 
}{}
\makeatletter
\@ifundefined{KOMAClassName}{
  \IfFileExists{parskip.sty}{%
    \usepackage{parskip}
  }{
    \setlength{\parindent}{0pt}
    \setlength{\parskip}{6pt plus 2pt minus 1pt}}
}{
  \KOMAoptions{parskip=half}}
\makeatother
\usepackage{xcolor}
\IfFileExists{xurl.sty}{\usepackage{xurl}}{} 
\IfFileExists{bookmark.sty}{\usepackage{bookmark}}{\usepackage{hyperref}}
\hypersetup{
  pdftitle={Variance-based global sensitivity analysis of numerical models using R},
  colorlinks=true,
  linkcolor={Maroon},
  filecolor={Maroon},
  citecolor={green},
  urlcolor={Blue},
  pdfcreator={LaTeX via pandoc}}
\usepackage[margin=1in]{geometry}
\usepackage{color}
\usepackage{fancyvrb}

\DefineVerbatimEnvironment{Highlighting}{Verbatim}{commandchars=\\\{\}}
\usepackage{framed}
\definecolor{shadecolor}{RGB}{248,248,248}
\newenvironment{Shaded}{\begin{snugshade}}{\end{snugshade}}

\newcommand{\AttributeTok}[1]{\textcolor[rgb]{0.77,0.63,0.00}{#1}}

\newcommand{\CommentTok}[1]{\textcolor[rgb]{0.56,0.35,0.01}{\textit{#1}}}

\newcommand{\ConstantTok}[1]{\textcolor[rgb]{0.00,0.00,0.00}{#1}}
\newcommand{\ControlFlowTok}[1]{\textcolor[rgb]{0.13,0.29,0.53}{\textbf{#1}}}

\newcommand{\DecValTok}[1]{\textcolor[rgb]{0.00,0.00,0.81}{#1}}

\newcommand{\ErrorTok}[1]{\textcolor[rgb]{0.64,0.00,0.00}{\textbf{#1}}}

\newcommand{\FloatTok}[1]{\textcolor[rgb]{0.00,0.00,0.81}{#1}}
\newcommand{\FunctionTok}[1]{\textcolor[rgb]{0.00,0.00,0.00}{#1}}

\newcommand{\NormalTok}[1]{#1}

\newcommand{\OtherTok}[1]{\textcolor[rgb]{0.56,0.35,0.01}{#1}}

\newcommand{\SpecialCharTok}[1]{\textcolor[rgb]{0.00,0.00,0.00}{#1}}

\newcommand{\StringTok}[1]{\textcolor[rgb]{0.31,0.60,0.02}{#1}}

\usepackage{longtable,booktabs,array}
\usepackage{calc} 
\usepackage{etoolbox}
\makeatletter
\patchcmd\longtable{\par}{\if@noskipsec\mbox{}\fi\par}{}{}
\makeatother
\IfFileExists{footnotehyper.sty}{\usepackage{footnotehyper}}{\usepackage{footnote}}
\makesavenoteenv{longtable}
\usepackage{graphicx}
\makeatletter
\def\maxwidth{\ifdim\Gin@nat@width>\linewidth\linewidth\else\Gin@nat@width\fi}
\def\maxheight{\ifdim\Gin@nat@height>\textheight\textheight\else\Gin@nat@height\fi}
\makeatother
\setkeys{Gin}{width=\maxwidth,height=\maxheight,keepaspectratio}
\makeatletter
\def\fps@figure{htbp}
\makeatother
\newlength{\cslhangindent}
\newlength{\csllabelwidth}
\newlength{\cslentryspacingunit} 
\newenvironment{CSLReferences}[2] 
 {
  \setlength{\parindent}{0pt}
  \ifodd #1
  \let\oldpar\par
  \def\par{\hangindent=\cslhangindent\oldpar}
  \fi
  \setlength{\parskip}{#2\cslentryspacingunit}
 }%
 {}
\usepackage{calc}

\usepackage{bm}
\ifLuaTeX
  \usepackage{selnolig}  
\fi

\title{Variance-based global sensitivity analysis of numerical models using R}
\author{Hossein Mohammadi\footnote{\href{mailto:h.mohammadi@exeter.ac.uk}{\nolinkurl{h.mohammadi@exeter.ac.uk}}} \(^1\), Peter Challenor\(^1\), and Cl\'ementine Prieur\(^2\)\\
\(^1\)College of Engineering, Mathematics and Physical Sciences, University of Exeter, UK\\
\(^2\)Univiversity of Grenoble Alpes, CNRS, INRIA, Grenoble, France}
\date{22, June, 2022}

\begin{document}
\maketitle
\begin{abstract}
Sensitivity analysis plays an important role in the development of computer models/simulators through identifying the contribution of each (uncertain) input factor to the model output variability. This report investigates different aspects of the variance-based global sensitivity analysis in the context of complex black-box computer codes. The analysis is mainly conducted using two \textbf{R} packages, namely \textbf{sensobol} (\protect\hyperlink{ref-puy2021}{Puy et al., 2021}) and \textbf{sensitivity} (\protect\hyperlink{ref-sensitiviti2021}{Iooss et al., 2021}). While the package \textbf{sensitivity} is equipped with a rich set of methods to conduct sensitivity analysis, especially in the case of models with dependent inputs, the package \textbf{sensobol} offers a bunch of user-friendly tools for the visualisation purposes. Several illustrative examples are supplied that allow the user to learn both packages easily and benefit from their features.

\par

\end{abstract}

\newcommand{\bx}{\mathbf{x}}
\newcommand{\by}{\mathbf{y}}
\newcommand{\bX}{\mathbf{X}}
\newcommand{\bcX}{\bm{\mathcal{X}}}
\newcommand{\bcY}{\bm{\mathcal{Y}}}
\newcommand{\bA}{\mathbf{A}}
\newcommand{\bB}{\mathbf{B}}
\newcommand{\V}{\mathbb{V}}
\newcommand{\E}{\mathbb{E}}
\newcommand{\Cov}{\mathbb{C}\text{ov}}
\newcommand{\diag}{\mathop{\mathrm{diag}}}

\hypertarget{intro}{%
\section{Introduction}\label{intro}}

In many situations it is impossible to implement physical experiments
due to its huge cost (time and/or monetary). To overcome this issue,
computer codes are developed in various scientific disciplines to
reproduce a physical mechanism relying on complex mathematical equations,
e.g., systems of nonlinear (partial) differential equations. Typically,
a computer model takes a set of input parameters/factors and produces
some output(s) quantity of interest. The inputs to the model are subject
to uncertainty due to various reasons such as our lack of knowledge about the real system,
missing physics, simplifying assumptions, measurement error, etc. The
input uncertainty results in the response uncertainty which is a measure
of model accuracy. Sensitivity analysis (SA) is a powerful technique
whereby we can understand the impact of each input on the variability of
model outputs.

In mathematical modelling, the main application of SA is to identify the
relative importance of each input factor based on its contribution to
the output variance. Such analysis is referred to as
\emph{factor prioritization} in the literature (\protect\hyperlink{ref-saltelli2004}{Andrea Saltelli et al., 2004}).
Moreover, SA is an integral part of uncertainty quantification which is
concerned with the estimation of uncertainty propagating through complex
models. In this paradigm, SA can help to decrease the output variance
by reducing the uncertainty in the inputs using techniques such as
\emph{factor screening} (\protect\hyperlink{ref-kleijnen2009-1}{Kleijnen, 2009}) or \emph{factor fixing}
(\protect\hyperlink{ref-saltelli2004}{Andrea Saltelli et al., 2004}). Factor screening allows us to eliminate insignificant
factors, especially when the model is data-driven and the number of
inputs exceeds the number of model evaluations (\protect\hyperlink{ref-song2016}{Song et al., 2016}). In factor fixing,
uncertainty of noninfluential inputs is ignored by fixing them at their
nominal values without a remarkable loss of information. Another approach is to consider noninfluential parameters as noise in the system.
Factor fixing is more suitable for physics-based codes where excluding
some of their input parameters is impractical.

There have been many methods to conduct SA, see e.g. (\protect\hyperlink{ref-borgonovo2016}{Borgonovo \& Plischke, 2016}; \protect\hyperlink{ref-daveiga2021}{Da Veiga et al., 2021}; \protect\hyperlink{ref-razavi2021}{Razavi et al., 2021}) for a comprehensive review of them.
Generally, they can be divided into two groups:

\begin{itemize}
\item
  \emph{Local SA} where the impact of inputs' variation on the output is
  assessed at a specific point in the parameter space. This is
  typically performed using partial derivatives (\protect\hyperlink{ref-gan2014}{Gan et al., 2014}). When
  computing the partial derivative with respect to a certain input,
  all other factors are held constant. As a result, the interaction
  among the inputs is neglected in local SA and the sensitivity
  measures are not comprehensive. However, local approaches
  are cheap in terms of computational cost.
\item
  \emph{Global SA (GSA)} where the output uncertainty is analysed over the
  entire variation range of the factors. Global methods provide more
  information about the model than local ones. For example, the
  interaction among the inputs can be captured in GSA since they are
  varied simultaneously. However, conducting GSA requires a large
  number of simulation runs which may not be affordable if the model
  is computationally intensive. A common way to combat this issue is
  to replace the simulator by a surrogate model (\protect\hyperlink{ref-marrel2012}{Marrel et al., 2012}; \protect\hyperlink{ref-sudret2008}{Sudret, 2008}). In Section \ref{stochastic-model}, a
  Gaussian process emulator (\protect\hyperlink{ref-GPML}{Rasmussen \& Williams, 2005}) serves as a surrogate to alleviate
  computational burden of applying SA to stochastic simulators. The focus of this work is on \emph{variance-based} GSA methods which are one of the most common way of measuring global sensitivity, even if the literature beyond variance-based SA is expanding very fast (see e.g., Chapter 6 in the book (\protect\hyperlink{ref-daveiga2021}{Da Veiga et al., 2021})). From now on any reference to SA refers to GSA.
\end{itemize}

This report covers various aspects of variance-based SA in the context of complex
black-box simulators from a practical point of view. The analysis is
carried out using \textbf{R} packages, namely \textbf{sensobol} (\protect\hyperlink{ref-puy2021}{Puy et al., 2021}) and
\textbf{sensitivity} (\protect\hyperlink{ref-sensitiviti2021}{Iooss et al., 2021}). These two libraries together offer a
comprehensive set of tools for performing SA. Several examples are
provided for each package that allow the user to learn them easily. A
taxonomy of available software packages for SA in programming languages
other than \textbf{R} can be found in (\protect\hyperlink{ref-qian2020}{G. Qian \& Mahdi, 2020}). The rest of the paper is
organised as follows. The statistical background of SA is reviewed in
Section \ref{sensitivity-indices}. Section \ref{SA-with-R} describes
how to conduct SA with the packages \textbf{sensobol} and \textbf{sensitivity}.
Section \ref{Shapley-effect} deals with the problem of applying SA to
simulators with dependent input variables. In Section
\ref{stochastic-model}, SA of stochastic model is discussed where
different outputs are observed for an identical input. Finally, the
conclusion is provided in Section \ref{conclusion}.

\hypertarget{sensitivity-indices}{%
\section{Sensitivity indices}\label{sensitivity-indices}}

Let us start this section by introducing computer models rigorously.
Suppose that the output of a deterministic numerical simulator is
governed by an unknown function \(f: \mathbb{R}^d \mapsto \mathbb{R}\)
with inputs \(\mathbf{x}= (x_1, \ldots, x_d)^\top\). Since the ``true''
value of the input parameters is unknown, each factor is considered as a
random variable and the associated uncertainty is described in terms of
probability distributions. This makes the model output a random variable
even if \(f\) is deterministic because the input uncertainty induces the
response uncertainty. As per convention that random variables are
denoted by capital letters, the model output is written as \[
Y = f(\mathbf{X}), ~\mathbf{X}= \left(X_1, \ldots, X_d \right)^\top ,
\] where \(\mathbf{X}\) consists of \(d\) statistically independent random
variables with known distributions. A variance-based SA technique
provides a framework whereby the variance of \(Y\) can be apportioned
into different sources of uncertainty in the inputs (\protect\hyperlink{ref-saltelli2004}{Andrea Saltelli et al., 2004}). One
of the most commonly used variance-based approaches is introduced by
Sobol' (\protect\hyperlink{ref-sobol2001}{Sobol', 2001}) relying on a functional decomposition of \(f\) as described below.

\hypertarget{sobol-indices}{%
\subsection{Sobol' indices}\label{sobol-indices}}

The Sobol' method (\protect\hyperlink{ref-sobol2001}{Sobol', 2001}) is a classical way of doing SA and has been successfully
employed in various application areas; see e.g., (\protect\hyperlink{ref-harenberg2019}{Harenberg et al., 2019}; \protect\hyperlink{ref-pianosi2016}{Pianosi et al., 2016}; \protect\hyperlink{ref-zhang2013}{Zhang et al., 2013}). The Sobol' sensitivity indices benefit from
several advantages including accuracy, clear interpretation and
straightforward implementation (\protect\hyperlink{ref-burnaev2017}{Burnaev et al., 2017}). The Sobol' method relies
on the following functional ANOVA (FANOVA) decomposition scheme
(\protect\hyperlink{ref-efron1981}{Efron \& Stein, 1981}; \protect\hyperlink{ref-sobol2001}{Sobol', 2001}) \begin{equation}
Y = f(\mathbf{X}) = f_0 + \sum_{i = 1}^d f_i(X_i) + \sum_{i}\sum_{j > i}f_{ij}(X_i, X_j) + \ldots + f_{1,2,\ldots,d}(\mathbf{X}) ,
\label{eq:fun-decompose}
\end{equation} wherein \(f_0\) is a constant. The remaining \(2^d - 1\)
elementary functions are centred (mean zero) and orthogonal (mutually
uncorrelated) with each other: \begin{align}
&\mathbb{E}\lbrack f_I(X_I) \rbrack = 0, \\
&\mathbb{E}\lbrack f_I(X_I) f_J(X_J) \rbrack = 0, ~\forall I \ne J ,
\end{align} in which
\(I, J \subseteq \mathcal{D} = \lbrace 1, \ldots, d \rbrace\). Applying
the variance operator, \(\mathbb{V}(\cdot)\), to the both sides of
\eqref{eq:fun-decompose} yields \begin{equation}
\mathbb{V}(Y) = \sum_{i = 1}^d\mathbb{V}_i + \sum_{i}\sum_{j > i}\mathbb{V}_{ij} + \ldots + \mathbb{V}_{1,2,\ldots,d} ,
\label{eq:var-decompose}
\end{equation} such that \begin{align}
&\mathbb{V}_i = \mathbb{V}\left(f_i(X_i)\right) = \mathbb{V}_{X_i} \left(\mathbb{E}_{\mathbf{X}_{\sim i}} \lbrack Y \mid X_i\rbrack \right), \\
&\mathbb{V}_{ij} = \mathbb{V}\left(f_{ij}(X_i, X_j)\right) = \mathbb{V}_{X_i, X_j} \left(\mathbb{E}_{\mathbf{X}_{\sim i, j}} \lbrack Y \mid X_i, X_j\rbrack \right) - \mathbb{V}_i - \mathbb{V}_j ,
\end{align} and the other terms are defined in a similar fashion. The
notation \(\mathbf{X}_{\sim i}\) (\(\mathbf{X}_{\sim i, j}\)) is used to
indicate all input factors except \(X_i\) (\(X_i\) and \(X_j\)).

The Sobol' indices are defined as \begin{equation}
S_i = \frac{\mathbb{V}_i}{\mathbb{V}(Y)}, ~ S_{ij} = \frac{\mathbb{V}_{ij}}{\mathbb{V}(Y)}, \ldots, 
\label{eq:sobol-ind}
\end{equation} where \(S_i\) is the first order (or main) effect of \(X_i\),
\(S_{ij}\) is the second order effect of \((X_i, X_j)\) (which represents
the contribution of interaction between \(X_i\) and \(X_j\) on the model
output uncertainty without their individual effects), and so on. Note
that the sum of all the sensitivity indices is equal to one:
\begin{equation}
\sum_{i = 1}^d S_i + \sum_{i}\sum_{j > i}S_{ij} + \ldots + S_{1,2,\ldots,d} = 1 .
\end{equation}

Among the terms defined in \eqref{eq:sobol-ind}, \(S_i\) is of great
significance; it reflects the direct contribution of \(X_i\) on the total
variance \(\mathbb{V}(Y)\), and is used as a measure of importance of
\(X_i\). The first order index can be interpreted as the expected
reduction in the total variance \(\mathbb{V}(Y)\) when \(X_i\) is fixed to a
constant (\protect\hyperlink{ref-saltelli2004}{Andrea Saltelli et al., 2004}). This is shown below using the law of total
variance \begin{equation}
S_i = \frac{\mathbb{V}_{X_i} \left(\mathbb{E}_{\mathbf{X}_{\sim i}} \lbrack Y \mid X_i\rbrack \right)}{\mathbb{V}(Y)} = \frac{\mathbb{V}(Y) - \mathbb{E}_{X_i} \lbrack \mathbb{V}_{\mathbf{X}_{\sim i}} \left( Y \mid X_i\right) \rbrack}{\mathbb{V}(Y)} . 
\end{equation} Another consequential sensitivity metric which
complements the first order effect is the total order index denoted by
\(S_{T_i}\). It measures the main effect of \(X_i\) together with its higher
order effects (interactions) with all the other factors. For example, in
a model with three input parameters, the total effect of \(X_1\) obeys:
\begin{equation}
S_{T_1} = S_1 + S_{1,2} + S_{1,3} + S_{1,2,3} . 
\end{equation} Notice that the total order index can be computed
directly \begin{equation}
S_{T_i} = \frac{\mathbb{V}(Y) - \mathbb{V}_{\mathbf{X}_{\sim i}} \left(\mathbb{E}_{X_i} \left[Y \mid \mathbf{X}_{\sim i} \right] \right)}{\mathbb{V}\left(Y \right)} = \frac{\mathbb{E}_{\mathbf{X}_{\sim i}} \left[ \mathbb{V}_{X_i} \left(Y \mid \mathbf{X}_{\sim i} \right) \right]}{\mathbb{V}(Y)} , 
\label{eq:total-effect}
\end{equation} in which
\(\mathbb{V}_{\mathbf{X}_{\sim i}} \left(\mathbb{E}_{X_i} \left[Y \mid \mathbf{X}_{\sim i} \right] \right)\)
stands for the first order effect of \(\mathbf{X}_{\sim i}\), i.e., all
factors but \(X_i\). Therefore, \(\mathbb{V}(Y)\) minus
\(\mathbb{V}_{\mathbf{X}_{\sim i}} \left(\mathbb{E}_{X_i} \left[Y \mid \mathbf{X}_{\sim i} \right] \right)\)
incorporates the contribution of all the terms that include \(X_i\) to the
output uncertainty.

In practice, only the first and total indices are considered for
sensitivity studies. The main effects are typically used for factor
prioritization, i.e., ranking the input parameters according to their
contribution to the total response variance. Also, the main effects can
be employed to identify \emph{additive} models where there is no
interaction between the factors. In this situation, we have
\(\sum_{i = 1}^{d}S_i = 1\). We note that the interaction of \(X_i\) with
the other factors is simply the difference between its main and total
indices: \(S_{T_i} - S_i\). The total effects are suit for factor fixing
where insignificant inputs are set to a given value over their range of
uncertainty (\protect\hyperlink{ref-saltelli2004}{Andrea Saltelli et al., 2004}). A factor \(X_i\) is said to be noninfluential
if its total order index is close to zero. Finally, the following
relation holds between the first and total effect indices given that the
input parameters are independent: \begin{equation}
\sum_{i = 1}^{d}S_i \le 1 \le \sum_{i = 1}^{d}S_{T_i} ,
\label{eq:first-total-relation}
\end{equation} which holds with equalities if the model is perfectly
additive (\protect\hyperlink{ref-song2016}{Song et al., 2016}).

\hypertarget{estimation-of-sobol-indices}{%
\subsection{Estimation of Sobol' indices}\label{estimation-of-sobol-indices}}

To calculate the sensitivity indices, the expectation and variance
operators (see e.g., Equation \eqref{eq:total-effect}) need to be
expressed in their integral forms. On the one hand, solving such
integrals numerically requires a huge number of model evaluations which
can be computationally expensive. On the other hand, it may not be
possible to find closed-form expressions for the decomposition
components in Equation \eqref{eq:fun-decompose}. To overcome these issues, several sampling-based estimators
(specially for the main and total effects) are developed which are
computationally ``efficient''. For example, Jansen (\protect\hyperlink{ref-jansen1999}{Jansen, 1999}) proposed
the following estimators for the first and total order indices
\begin{align}
\label{eq:jansen-main}
&\hat{S}_i = 1 - \frac{\frac{1}{2N} \sum_{i = 1}^{d} \left\lbrack f(\mathbf{B}) - f \left(\mathbf{A}^{(i)}_{B} \right) \right\rbrack^2 }{\mathbb{V}\left(Y \right)} , \\
\label{eq:jansen-total}
&\hat{S}_{T_i} = \frac{\frac{1}{2N} \sum_{i = 1}^{d} \left\lbrack f(\mathbf{A}) - f \left(\mathbf{A}^{(i)}_{B} \right) \right\rbrack^2 }{\mathbb{V}\left(Y \right)} .
\end{align}
Here, \(N\) is the number of samples and \(\mathbf{A}\) and
\(\mathbf{B}\) are two random matrices of size \(N\times d\). The matrix
\(\mathbf{A}^{(i)}_{B}\) is the same as \(\mathbf{A}\) except that its
\(i\)-th column comes from \(\mathbf{B}\); see below. \[
\mathbf{A}= \begin{bmatrix} 
    a_{11} & \dots  & a_{1d} \\
    \vdots & \ddots & \vdots \\
    a_{N1} & \dots  & a_{Nd} 
    \end{bmatrix} , ~ \mathbf{B}= \begin{bmatrix} 
    b_{11} & \dots  & b_{1d} \\
    \vdots & \ddots & \vdots \\
    b_{N1} & \dots  & b_{Nd} 
    \end{bmatrix} ~\Rightarrow
\mathbf{A}^{(i)}_{B} = \begin{bmatrix} 
    a_{11} & \dots & b_{1i} & \dots & a_{1d} \\
    \vdots & \dots & \vdots & \dots & \vdots \\
    a_{N1} & \dots & b_{Ni} & \dots & a_{Nd}
    \end{bmatrix}.
\] Note that the matrices \(\mathbf{A}\), \(\mathbf{B}\) and
\(\mathbf{A}^{(i)}_{B}, i = 1, \ldots, d,\) together provide a total
number of \(N\times(d + 2)\) samples to be evaluated by the simulator. This way
of generating samples is referred to as the \emph{pick-freeze} scheme first
introduced by Sobol' (\protect\hyperlink{ref-sobol2001}{Sobol', 2001}). The
pick-freeze scheme is used in different estimators which can be found in e.g., (\protect\hyperlink{ref-homma1996}{Homma \& Saltelli, 1996}; \protect\hyperlink{ref-janon2014}{Janon et al., 2014}; \protect\hyperlink{ref-saltelli2002}{Andrea Saltelli, 2002}; \protect\hyperlink{ref-sobol2001}{Sobol', 2001}). An alternative to sampling-based methods for the estimation of Sobol' indices is \emph{spectral} approaches (\protect\hyperlink{ref-cukier1978}{Cukier et al., 1978}; \protect\hyperlink{ref-saltelli1999}{A. Saltelli et al., 1999}). They are based on
a spectral decomposition of \(f\) with some regularity assumptions in terms of decay conditions on the coefficients. Such spectral approaches are beyond the scope of this work.

The matrices \(A\) and \(B\) used in the pick-freeze scheme can be built
using a Monte Carlo (MC) approach. It is found, however, that the MC methods are not
efficient for the estimation of the Sobol' indices specially when \(N\) is small (\protect\hyperlink{ref-daveiga2021}{Da Veiga et al., 2021}). Besides, a clustering of points can happen in some regions of the space since the MC approach does not record the history
of previous points. Quasi-MC (QMC) methods such as Halton (\protect\hyperlink{ref-halton1960}{Halton, 1960}) or Sobol'
(\protect\hyperlink{ref-sobol1967}{Sobol', 1967}) sequences yield a more uniform spread of points than MC ones. A QMC
method uses a low \emph{discrepancy} sequence to produce samples where the
discrepancy criterion is a measure of deviation from a perfectly uniform
distribution of points (\protect\hyperlink{ref-niederreiter1992}{Niederreiter, 1992}). The
\(\mathbb{L}^2\)-discrepancy of the set \(A \subset [0, 1]^d\) is defined as
\begin{equation}
    D_2\left(A\right) = \left\lbrack \int_{[0, 1]^d} \left\lvert \frac{ \text{Num}\left(A, h_{\bx} \right)}{N} - \text{Vol} (h_{\bx}) \right\rvert^2 \, d\bx \right\rbrack^{1/2} , 
\end{equation} where \(h_{\bx}\) denotes the interval
\([0, \bx) = [0, x_1) \times [0, x_2)\times\ldots \times [0, x_d)\),
\(\text{Num} \left(A, h_{\bx} \right)\) is the number of points of \(A\)
falling in \(h_{\bx}\), and \(\text{Vol}(h_{\bx})\) denotes the volume of
\(h_{\bx}\). In this work, the samples are generated via the Sobol' QMC method (\protect\hyperlink{ref-sobol1967}{Sobol', 1967}).

\begin{figure}

{\centering \includegraphics[width=0.32\linewidth]{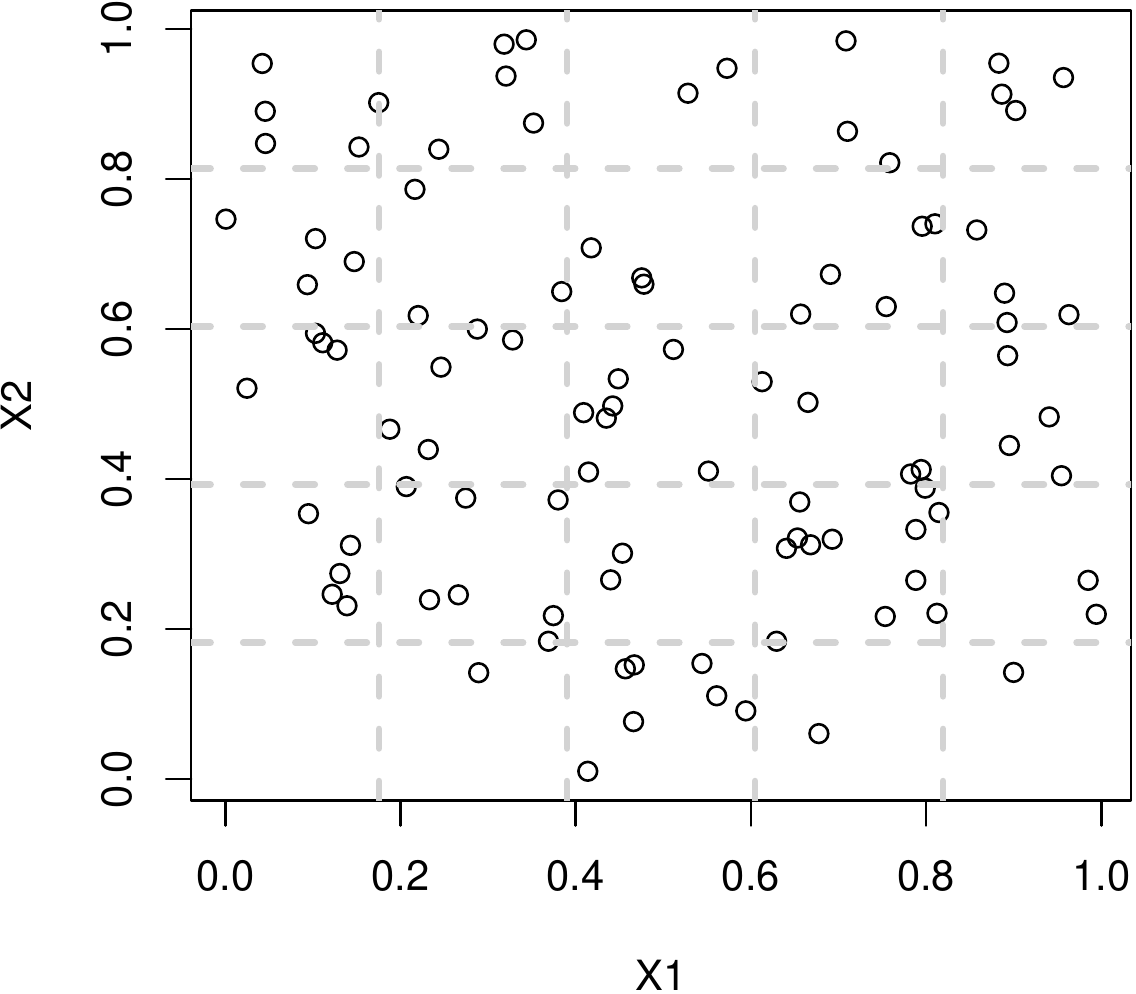} \includegraphics[width=0.32\linewidth]{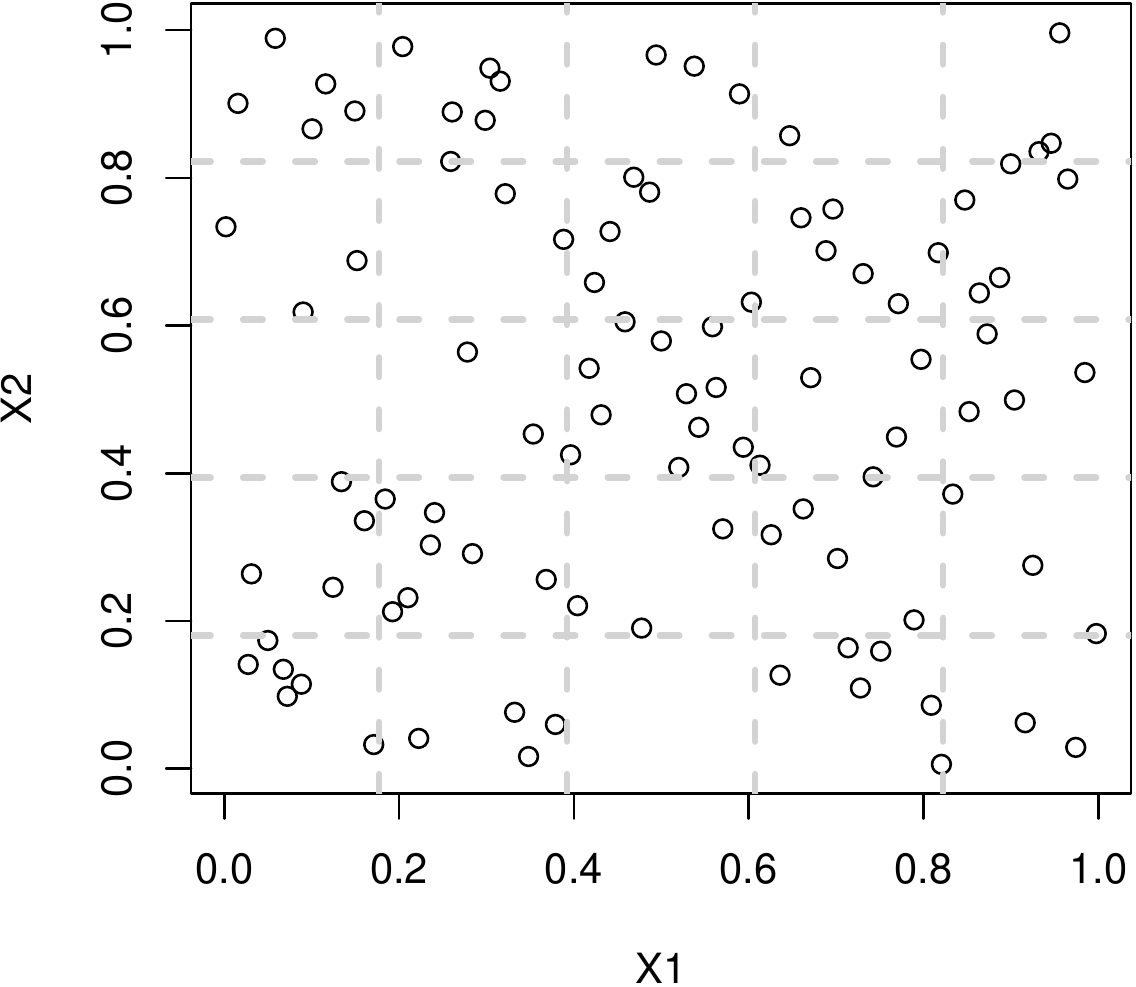} \includegraphics[width=0.32\linewidth]{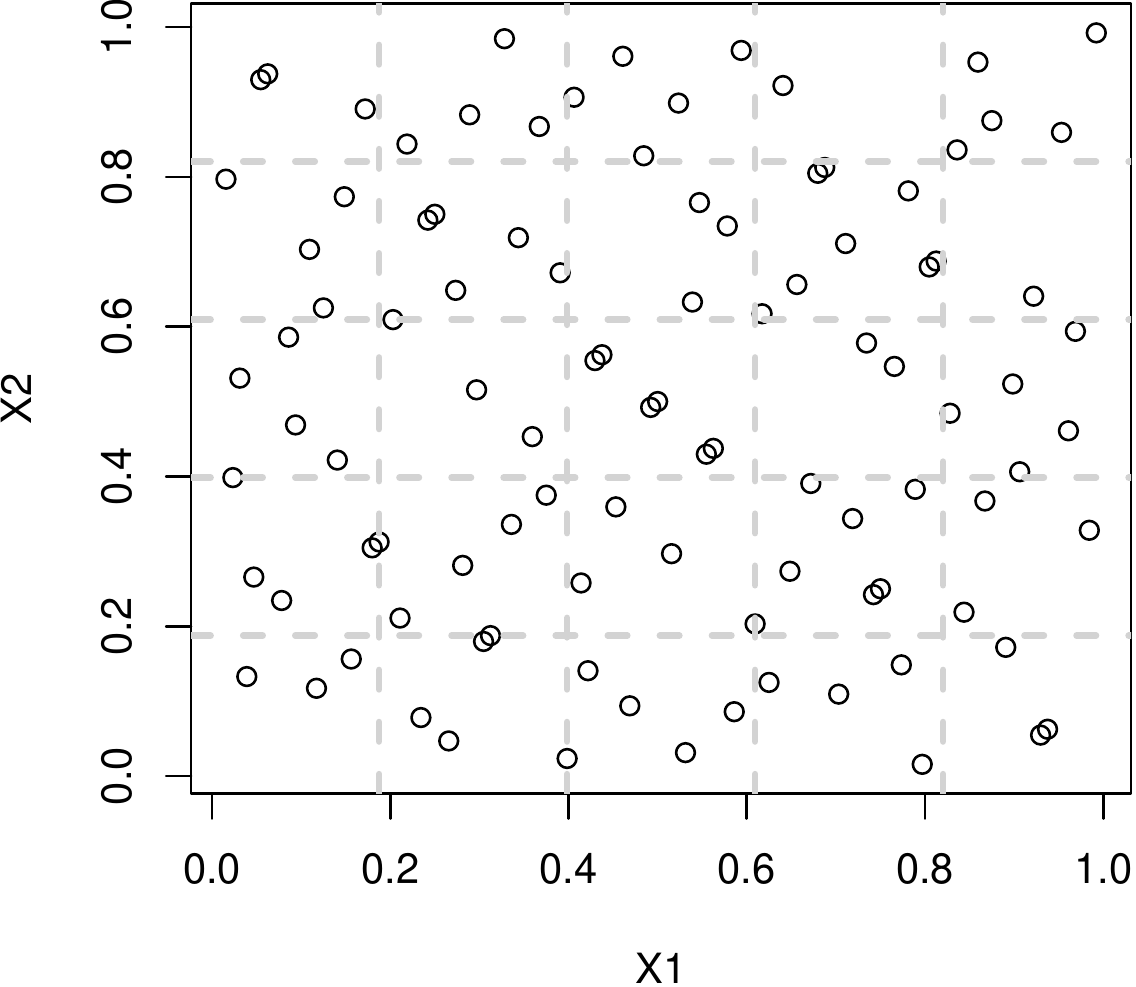} 

}

\caption{Sample points generated by MC (left), LHS (middle) and QMC (right) in $\lbrack 0, 1\rbrack ^2$. The LHS and QMC samples are produced using the functions \texttt{lhsDesign()} and \texttt{sobol()} implemented in the packages \textbf{DiceDesign} and \textbf{randtoolbox}, respectively. The number of samples is: $N = 100$.}\label{fig:MC-sampling}
\end{figure}

A Latin hypercube sampling (LHS) design (\protect\hyperlink{ref-mckay1979}{McKay et al., 1979}; \protect\hyperlink{ref-stein1987}{Stein, 1987}) is another strategy to place the sample points in a uniform manner across the input space. The iterative construction of LHS is
discussed in (\protect\hyperlink{ref-gilquin2017}{Gilquin et al., 2017}; \protect\hyperlink{ref-qian2009}{P. Z. G. Qian, 2009}). (\protect\hyperlink{ref-gilquin2019}{Gilquin et al., 2019}; \protect\hyperlink{ref-tissot2015}{Tissot \& Prieur, 2015}) proposed a \emph{replicated} LHS for the estimation of Sobol' indices. Figure \ref{fig:MC-sampling} illustrates the samples based on MC (left), LHS (middle) and QMC (right). The two random variables follow a uniform distribution in \([0, 1]^2\), and \(N = 100\). The QMC and LHS samples are obtained by
the packages \textbf{randtoolbox} (\protect\hyperlink{ref-randtoolbox2020}{Chalabi et al., 2020}) and \textbf{DiceDesign}
(\protect\hyperlink{ref-DiceDesign}{Dupuy et al., 2015}), respectively.

\begin{Shaded}
\begin{Highlighting}[]
\FunctionTok{library}\NormalTok{(randtoolbox)}
\FunctionTok{library}\NormalTok{(DiceDesign)}
\FunctionTok{set.seed}\NormalTok{(}\DecValTok{123}\NormalTok{)}
\NormalTok{d }\OtherTok{\textless{}{-}} \DecValTok{2}
\NormalTok{N }\OtherTok{\textless{}{-}} \DecValTok{100}
\NormalTok{A\_MC }\OtherTok{\textless{}{-}} \FunctionTok{data.frame}\NormalTok{(}\FunctionTok{matrix}\NormalTok{(}\FunctionTok{runif}\NormalTok{(N}\SpecialCharTok{*}\NormalTok{d), }\AttributeTok{ncol =}\NormalTok{ d))}
\NormalTok{A\_LHS }\OtherTok{\textless{}{-}} \FunctionTok{data.frame}\NormalTok{(}\FunctionTok{lhsDesign}\NormalTok{(N, d)}\SpecialCharTok{$}\NormalTok{design)}
\NormalTok{A\_QMC }\OtherTok{\textless{}{-}} \FunctionTok{data.frame}\NormalTok{(}\FunctionTok{sobol}\NormalTok{(N, d))}
\FunctionTok{plot}\NormalTok{(A\_MC); }\FunctionTok{grid}\NormalTok{(}\AttributeTok{nx =} \DecValTok{5}\NormalTok{, }\AttributeTok{ny =} \DecValTok{5}\NormalTok{, }\AttributeTok{lty =} \DecValTok{2}\NormalTok{, }\AttributeTok{lwd =} \FloatTok{2.5}\NormalTok{)}
\FunctionTok{plot}\NormalTok{(A\_LHS); }\FunctionTok{grid}\NormalTok{(}\AttributeTok{nx =} \DecValTok{5}\NormalTok{, }\AttributeTok{ny =} \DecValTok{5}\NormalTok{, }\AttributeTok{lty =} \DecValTok{2}\NormalTok{, }\AttributeTok{lwd =} \FloatTok{2.5}\NormalTok{)}
\FunctionTok{plot}\NormalTok{(A\_QMC); }\FunctionTok{grid}\NormalTok{(}\AttributeTok{nx =} \DecValTok{5}\NormalTok{, }\AttributeTok{ny =} \DecValTok{5}\NormalTok{, }\AttributeTok{lty =} \DecValTok{2}\NormalTok{, }\AttributeTok{lwd =} \FloatTok{2.5}\NormalTok{)}
\end{Highlighting}
\end{Shaded}

\hypertarget{SA-with-R}{%
\section{\texorpdfstring{Sensitivity analysis with \textbf{R}}{Sensitivity analysis with R}}\label{SA-with-R}}

In this section, the packages \textbf{sensobol} (v1.0.1) (\protect\hyperlink{ref-puy2021}{Puy et al., 2021}) and
\textbf{sensitivity} (v1.25.0) (\protect\hyperlink{ref-sensitiviti2021}{Iooss et al., 2021}) are employed to conduct SA
on two test functions serve as ``true'' models. The functions are selected such
that their sensitivity indices (first and total) can be computed
analytically. This allows us to compare the estimated indices with their
actual values. The number of sample points and
bootstrap replicates to get confidence intervals for the estimates is
\(N = 5000\) and \(R = 1000\), respectively.

\hypertarget{polynomial-function}{%
\subsection{Polynomial function}\label{polynomial-function}}

The first test example is a polynomial function with four independent
input variables distributed uniformly \begin{equation}
Y =  3X_1^2 + X_2X_3 - 2X_4, \quad X_1, \ldots, X_4 \sim \mathcal{U}(0, 1) .
\label{eq:polynom-fun}
\end{equation} To compute the sensitivity indices analytically we need
the variance of each component in the function. The variance of the
first, the second and the last term on the right hand side of
\eqref{eq:polynom-fun} is\\
\begin{align}
\label{eq:first-term-var}
\mathbb{V}\left(3X_1^2 \right) &= 9\left(\mathbb{E}\lbrack X_1^4\rbrack - \mathbb{E}\lbrack X_1^2\rbrack^2\right) = 9\left(1/5 - (1/3)^2\right) = 0.8 , \\
\label{eq:second-term-var}
\mathbb{V}\left(X_2X_3 \right) &= \mathbb{E}\lbrack X_2^2\rbrack \mathbb{E}\lbrack X_3^2\rbrack - \mathbb{E}\lbrack X_2\rbrack^2 \mathbb{E}\lbrack X_3\rbrack^2 = 1/9 - 1/16 \approx 0.049 ,\\ 
\mathbb{V}\left(-2X_4\right) &\approx 0.333 ,
\label{eq:third-term-var}
\end{align} given that the \(q\)-th moment of \(X \sim \mathcal{U}(a, b)\)
has the following form \begin{equation*}
\mathbb{E}\lbrack X^q \rbrack = \frac{b^{q + 1} - a^{q + 1}}{(q + 1)(b - a)} .
\end{equation*}

The total output variance is simply the sum of all the variances in
Equations \eqref{eq:first-term-var}-\eqref{eq:third-term-var} since the
input variables are independent:
\(\mathbb{V}\left(Y \right) = 0.8 + 0.049 + 0.333 = 1.182\). Now, we can
easily calculate the theoretical sensitivity indices as described in
Section \ref{sensitivity-indices}. The results are summarised in Table
\ref{tab:polynomial-indices}.

\begin{table}
\centering
\caption{Analytical sensitivity indices for the polynomial function defined in $(\ref{eq:polynom-fun}).$} 
\begin{tabular}{l|c c c c}
  & First order && Total order \\
\hline
$X_1$ & 0.677 && 0.677 \\
$X_2$ & 0 && 0.041 \\
$X_3$ & 0 && 0.041 \\
$X_4$ & 0.282 && 0.282 \\
\end{tabular}
\label{tab:polynomial-indices}
\end{table}

The sensitivity indices of the polynomial function are estimated with the functions
\texttt{sobol\_indices()} (\textbf{sensobol} package) and \texttt{sobolSalt()}
(\textbf{sensitivity} package) using their default settings. This is shown in the code chunk below. To
create the random matrices \(\mathbf{A}\) and \(\mathbf{B}\) we use the
function \texttt{sobol\_matrices()} implemented in the package \textbf{sensobol}. The outcome of this
function (stored in the object \texttt{mat}) consists of \(\mathbf{A}\), \(\mathbf{B}\)
and \(\mathbf{A}^{(i)}_{B}, i = 1, \ldots, d\). By default, the function \texttt{sobol\_matrices()} generates samples based on the Sobol' QMC method via a call to the function \texttt{sobol()} of the \textbf{randtoolbox} package.
The results are rounded to the third significant digit. We observe that the results of the two packages are very similar and the estimated indices are close to their actual
values (Table \ref{tab:polynomial-indices}).

\begin{Shaded}
\begin{Highlighting}[]
\FunctionTok{library}\NormalTok{(sensobol)}
\FunctionTok{library}\NormalTok{(sensitivity)}
\FunctionTok{library}\NormalTok{(data.table)}
\NormalTok{fun }\OtherTok{\textless{}{-}} \ControlFlowTok{function}\NormalTok{(xx) \{}
\NormalTok{  yy }\OtherTok{\textless{}{-}} \DecValTok{3}\SpecialCharTok{*}\NormalTok{xx[, }\DecValTok{1}\NormalTok{]}\SpecialCharTok{\^{}}\DecValTok{2} \SpecialCharTok{+}\NormalTok{ xx[, }\DecValTok{2}\NormalTok{]}\SpecialCharTok{*}\NormalTok{xx[, }\DecValTok{3}\NormalTok{] }\SpecialCharTok{{-}} \DecValTok{2}\SpecialCharTok{*}\NormalTok{xx[, }\DecValTok{4}\NormalTok{]}
  \FunctionTok{return}\NormalTok{(yy)}
\NormalTok{\}}
\NormalTok{d }\OtherTok{\textless{}{-}} \DecValTok{4}
\NormalTok{N }\OtherTok{\textless{}{-}} \DecValTok{5000}
\NormalTok{R }\OtherTok{\textless{}{-}} \DecValTok{1000}
\NormalTok{params }\OtherTok{\textless{}{-}} \FunctionTok{paste}\NormalTok{(}\StringTok{"$X\_"}\NormalTok{, }\DecValTok{1}\SpecialCharTok{:}\NormalTok{d, }\StringTok{"$"}\NormalTok{, }\AttributeTok{sep =} \StringTok{""}\NormalTok{)}
\NormalTok{mat }\OtherTok{\textless{}{-}} \FunctionTok{sobol\_matrices}\NormalTok{(}\AttributeTok{N =}\NormalTok{ N, }\AttributeTok{params =}\NormalTok{ params)}
\NormalTok{Y }\OtherTok{\textless{}{-}} \FunctionTok{fun}\NormalTok{(mat)}
\NormalTok{sensobol\_ind }\OtherTok{\textless{}{-}} \FunctionTok{sobol\_indices}\NormalTok{(}\AttributeTok{Y =}\NormalTok{ Y, }\AttributeTok{N =}\NormalTok{ N, }\AttributeTok{params =}\NormalTok{ params, }
                              \AttributeTok{boot =} \ConstantTok{TRUE}\NormalTok{, }\AttributeTok{R =}\NormalTok{ R)}
\NormalTok{cols }\OtherTok{\textless{}{-}} \FunctionTok{colnames}\NormalTok{(sensobol\_ind}\SpecialCharTok{$}\NormalTok{results)[}\DecValTok{1}\SpecialCharTok{:}\DecValTok{5}\NormalTok{]}
\NormalTok{sensobol\_ind}\SpecialCharTok{$}\NormalTok{results[, (cols)}\SpecialCharTok{:}\ErrorTok{=} \FunctionTok{round}\NormalTok{(.SD, }\DecValTok{3}\NormalTok{), .SDcols }\OtherTok{=}\NormalTok{ (cols)]}
\FunctionTok{print}\NormalTok{(sensobol\_ind)}
\end{Highlighting}
\end{Shaded}

\begin{verbatim}
## 
## First-order estimator: saltelli | Total-order estimator: jansen 
## 
## Total number of model runs: 30000 
## 
## Sum of first order indices: 0.9938516 
##    original bias std.error low.ci high.ci sensitivity parameters
## 1:    0.677    0     0.016  0.645   0.709          Si      $X_1$
## 2:    0.018    0     0.003  0.011   0.024          Si      $X_2$
## 3:    0.018    0     0.003  0.012   0.024          Si      $X_3$
## 4:    0.282    0     0.010  0.263   0.301          Si      $X_4$
## 5:    0.677    0     0.012  0.653   0.701          Ti      $X_1$
## 6:    0.023    0     0.001  0.022   0.025          Ti      $X_2$
## 7:    0.023    0     0.001  0.022   0.025          Ti      $X_3$
## 8:    0.282    0     0.005  0.271   0.293          Ti      $X_4$
\end{verbatim}

\begin{Shaded}
\begin{Highlighting}[]
\NormalTok{sensitivity\_ind }\OtherTok{\textless{}{-}} \FunctionTok{sobolSalt}\NormalTok{(}\AttributeTok{model =}\NormalTok{ fun, }\AttributeTok{X1 =}\NormalTok{ mat[}\DecValTok{1}\SpecialCharTok{:}\NormalTok{N, ], }
                               \AttributeTok{X2 =}\NormalTok{ mat[(N}\SpecialCharTok{+}\DecValTok{1}\NormalTok{)}\SpecialCharTok{:}\NormalTok{(}\DecValTok{2}\SpecialCharTok{*}\NormalTok{N), ], }\AttributeTok{nboot =}\NormalTok{ R)}
\FunctionTok{print}\NormalTok{(}\FunctionTok{round}\NormalTok{(sensitivity\_ind}\SpecialCharTok{$}\NormalTok{S, }\DecValTok{3}\NormalTok{))  }\CommentTok{\# First order indices}
\end{Highlighting}
\end{Shaded}

\begin{verbatim}
##    original bias std. error min. c.i. max. c.i.
## X1    0.677    0      0.007     0.664     0.690
## X2    0.018    0      0.014    -0.007     0.047
## X3    0.017    0      0.014    -0.009     0.047
## X4    0.282    0      0.012     0.257     0.307
\end{verbatim}

\begin{Shaded}
\begin{Highlighting}[]
\FunctionTok{print}\NormalTok{(}\FunctionTok{round}\NormalTok{(sensitivity\_ind}\SpecialCharTok{$}\NormalTok{T, }\DecValTok{3}\NormalTok{))  }\CommentTok{\# Total order indices}
\end{Highlighting}
\end{Shaded}

\begin{verbatim}
##    original bias std. error min. c.i. max. c.i.
## X1    0.677    0      0.012     0.650     0.701
## X2    0.024    0      0.001     0.022     0.025
## X3    0.023    0      0.001     0.022     0.025
## X4    0.282    0      0.006     0.270     0.294
\end{verbatim}

\begin{figure}

{\centering \includegraphics[width=0.55\linewidth]{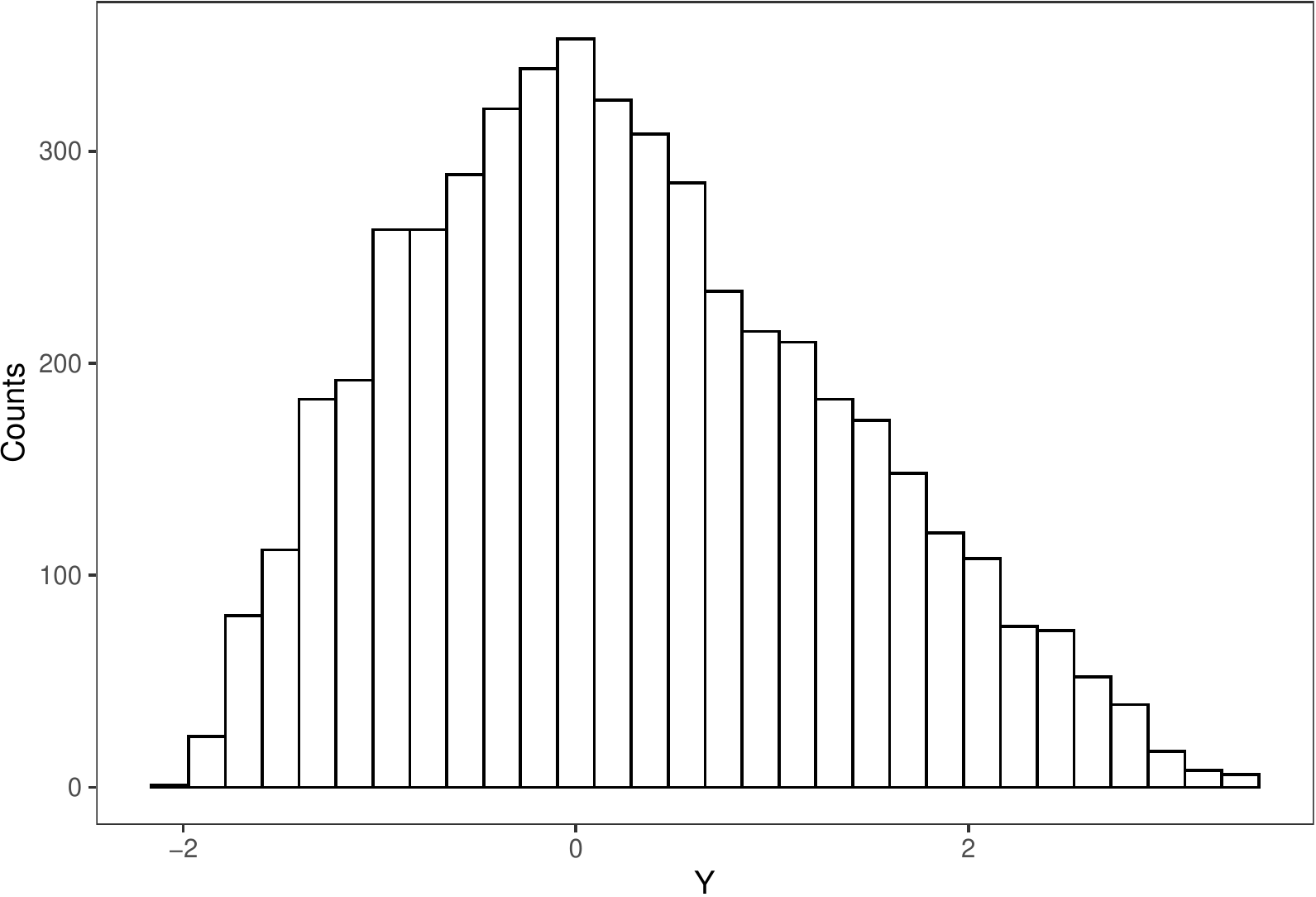} 

}

\caption{The empirical distribution of $Y$ for the polynomial function defined in (\ref{eq:polynom-fun}) using the function \texttt{plot\_uncertainty()} implemented in the \textbf{sensobol} package.}\label{fig:Y-hist}
\end{figure}

\begin{figure}

{\centering \includegraphics[width=1\linewidth]{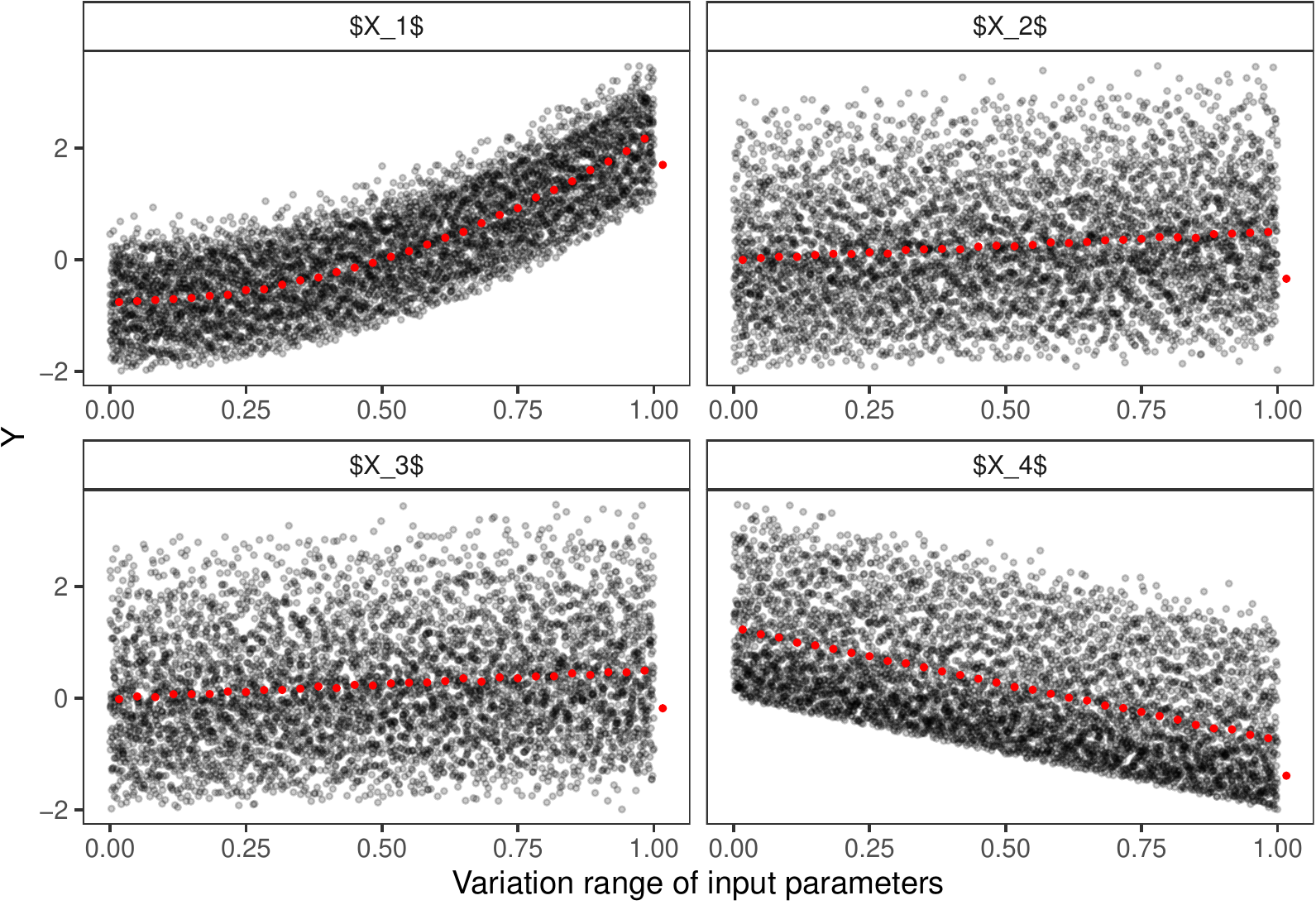} 

}

\caption{The scatter plots obtained by the function \texttt{plot\_scatter()} implemented in the \textbf{sensobol} package. The black dots are responses of the polynomial function given $X_i$. The domain of each factor is divided into bins and the red dots represent the mean of $Y \mid X_i$ in every bin.}\label{fig:polynomial-scatter-plot}
\end{figure}

It is worth mentioning
that the package \textbf{sensobol} offers several useful visualisation tools
such as \texttt{plot\_uncertainty()} and \texttt{plot\_scatter()} relying on the package
\textbf{ggplot2} (\protect\hyperlink{ref-ggplot2}{Wickham, 2016}). The former plots the histogram of the model
response and the latter gives an scatter plot against each input
parameter. Figures \ref{fig:Y-hist} and
\ref{fig:polynomial-scatter-plot} display the graphs obtained by
\texttt{plot\_uncertainty()} and \texttt{plot\_scatter()} for the polynomial function,
respectively. The function \texttt{plot\_scatter()} divides the domain of each
input \(X_i\) into bins and computes the mean of \(Y \mid X_i\) in every
bin. The red dots in Figure \ref{fig:polynomial-scatter-plot} represent
those means.

\begin{Shaded}
\begin{Highlighting}[]
\FunctionTok{library}\NormalTok{(ggplot2)}
\FunctionTok{plot\_uncertainty}\NormalTok{(}\AttributeTok{Y =}\NormalTok{ Y, }\AttributeTok{N =}\NormalTok{ N) }\SpecialCharTok{+} \FunctionTok{labs}\NormalTok{(}\AttributeTok{y =} \StringTok{"Counts"}\NormalTok{, }\AttributeTok{x =} \StringTok{"Y"}\NormalTok{) }
\end{Highlighting}
\end{Shaded}

\begin{Shaded}
\begin{Highlighting}[]
\FunctionTok{plot\_scatter}\NormalTok{(}\AttributeTok{data =}\NormalTok{ mat, }\AttributeTok{N =}\NormalTok{ N, }\AttributeTok{Y =}\NormalTok{ Y, }\AttributeTok{params =}\NormalTok{ params) }\SpecialCharTok{+} 
  \FunctionTok{labs}\NormalTok{(}\AttributeTok{y =} \StringTok{"Y"}\NormalTok{, }\AttributeTok{x =} \StringTok{"Variation range of input parameters"}\NormalTok{)}
\end{Highlighting}
\end{Shaded}

\hypertarget{ishigami-function}{%
\subsection{Ishigami function}\label{ishigami-function}}

\begin{figure}

{\centering \includegraphics[width=1\linewidth]{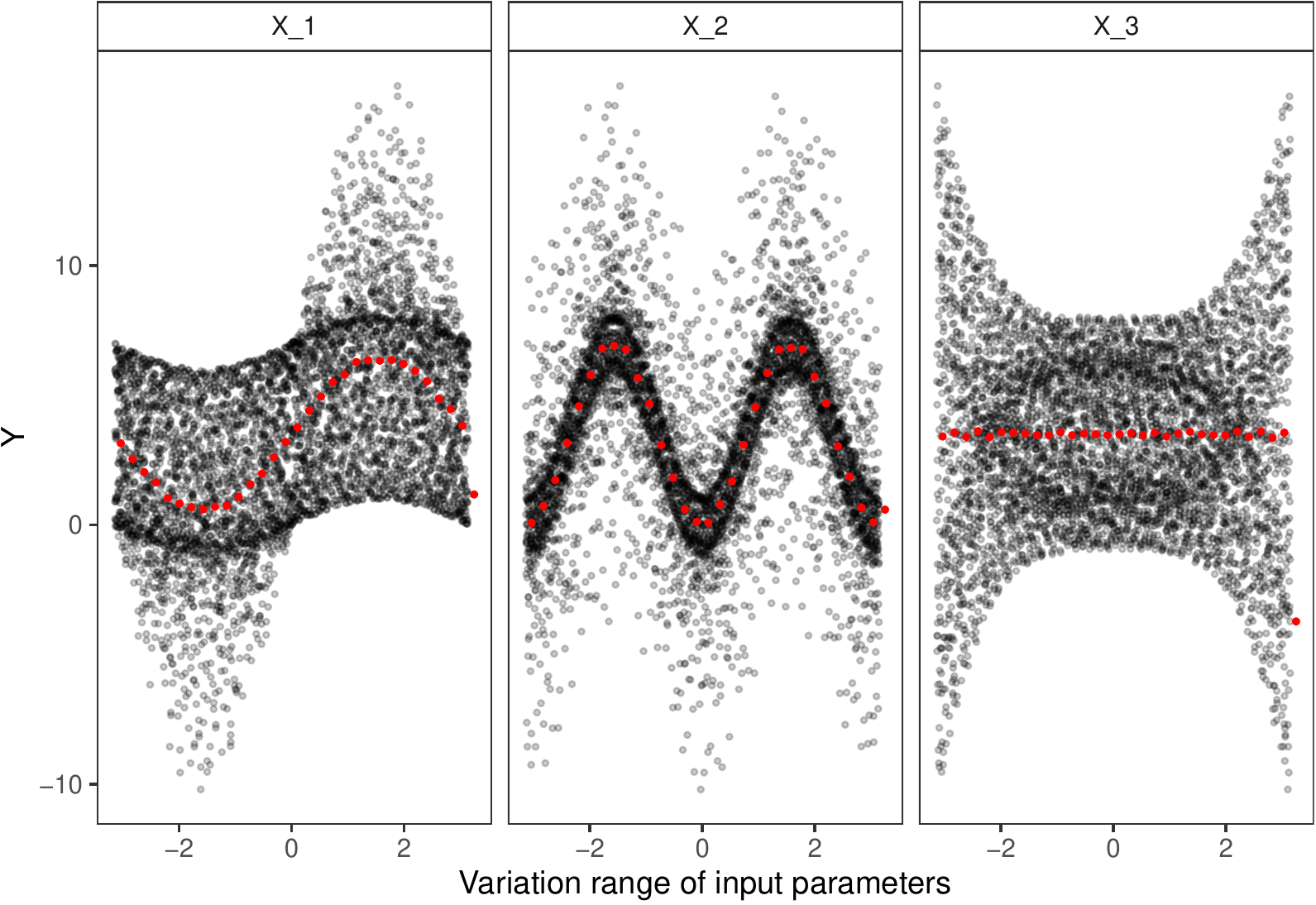} 

}

\caption{The scatter plots obtained by \texttt{plot\_scatter()} for the Ishigami function. The red dots are the mean of $Y \mid X_i$ in the bins created over the domain of $X_i$.}\label{fig:ishigami-scatter-plot}
\end{figure}

The Ishigami function (\protect\hyperlink{ref-ishigami1990}{Ishigami \& Homma, 1990}) is commonly used as a benchmark
example for sensitivity studies. It is a 3-dimensional, highly nonlinear
function expressed by \begin{equation}
Y = \sin(X_1) + 7\sin^2(X_2) + 0.1X_3^4\sin(X_1), \quad X_1, X_2, X_3\sim \mathcal{U}(-\pi, \pi) .
\label{eq:ishigami-fun}
\end{equation} The scatter plot of the Ishigami's input variables is
demonstrated in Figure \ref{fig:ishigami-scatter-plot} using the
function \texttt{plot\_scatter()}. It is observed that the main effect of \(X_3\)
(i.e., \(\mathbb{V}\left(\mathbb{E}\lbrack Y \mid X_3\rbrack \right)\)) is
possibly zero as the red dots (the mean of \(Y \mid X_i\) in the bins) has
a flat pattern. The theoretical Sobol' indices for the Ishigami function
are calculated in (\protect\hyperlink{ref-ishigami1990}{Ishigami \& Homma, 1990}; \protect\hyperlink{ref-sudret2008}{Sudret, 2008}) and are summarised in
Table \ref{tab:ishigami-indices}.

\begin{table}
\centering
\caption{Theoretical values of the first and total order sensitivity indices for the Ishigami function.} 
\begin{tabular}{l|c c c c}
  & First order && Total order \\
\hline
$X_1$ & 0.314 && 0.558 \\
$X_2$ & 0.442 && 0.442 \\
$X_3$ & 0 && 0.244 \\
\end{tabular}
\label{tab:ishigami-indices}
\end{table}

Now we use the two packages to estimate the sensitivity indices and
compare them with their actual values; see the code chunk below. The
Ishigami function is implemented in the package \textbf{sensitivity} and can
be called via \texttt{ishigami.fun()}. The results show that the two packages
have a good performance in estimating the sensitivity indices.

\begin{Shaded}
\begin{Highlighting}[]
\NormalTok{d }\OtherTok{\textless{}{-}} \DecValTok{3}
\NormalTok{params }\OtherTok{\textless{}{-}} \FunctionTok{paste}\NormalTok{(}\StringTok{"X\_"}\NormalTok{, }\DecValTok{1}\SpecialCharTok{:}\NormalTok{d, }\AttributeTok{sep =} \StringTok{""}\NormalTok{)}
\NormalTok{mat }\OtherTok{\textless{}{-}} \FunctionTok{sobol\_matrices}\NormalTok{(}\AttributeTok{N =}\NormalTok{ N, }\AttributeTok{params =}\NormalTok{ params)}
\NormalTok{mat }\OtherTok{\textless{}{-}} \DecValTok{2}\SpecialCharTok{*}\NormalTok{pi}\SpecialCharTok{*}\NormalTok{mat }\SpecialCharTok{{-}}\NormalTok{ pi }
\NormalTok{Y }\OtherTok{\textless{}{-}} \FunctionTok{ishigami.fun}\NormalTok{(mat)}
\NormalTok{sensobol\_ind }\OtherTok{\textless{}{-}} \FunctionTok{sobol\_indices}\NormalTok{(}\AttributeTok{Y =}\NormalTok{ Y, }\AttributeTok{N =}\NormalTok{ N, }\AttributeTok{params =}\NormalTok{ params, }
                              \AttributeTok{boot =} \ConstantTok{TRUE}\NormalTok{, }\AttributeTok{R =}\NormalTok{ R)}
\NormalTok{cols }\OtherTok{\textless{}{-}} \FunctionTok{colnames}\NormalTok{(sensobol\_ind}\SpecialCharTok{$}\NormalTok{results)[}\DecValTok{1}\SpecialCharTok{:}\DecValTok{5}\NormalTok{]}
\NormalTok{sensobol\_ind}\SpecialCharTok{$}\NormalTok{results[, (cols)}\SpecialCharTok{:}\ErrorTok{=} \FunctionTok{round}\NormalTok{(.SD, }\DecValTok{3}\NormalTok{), .SDcols }\OtherTok{=}\NormalTok{ (cols)]}
\FunctionTok{print}\NormalTok{(sensobol\_ind)}
\end{Highlighting}
\end{Shaded}

\begin{verbatim}
## 
## First-order estimator: saltelli | Total-order estimator: jansen 
## 
## Total number of model runs: 25000 
## 
## Sum of first order indices: 0.7677866 
##    original  bias std.error low.ci high.ci sensitivity parameters
## 1:    0.316 0.001     0.021  0.274   0.355          Si        X_1
## 2:    0.446 0.000     0.017  0.412   0.480          Si        X_2
## 3:    0.007 0.000     0.017 -0.026   0.039          Si        X_3
## 4:    0.561 0.001     0.020  0.520   0.599          Ti        X_1
## 5:    0.444 0.000     0.010  0.425   0.462          Ti        X_2
## 6:    0.244 0.000     0.006  0.232   0.257          Ti        X_3
\end{verbatim}

\begin{Shaded}
\begin{Highlighting}[]
\NormalTok{sensitivity\_ind }\OtherTok{\textless{}{-}} \FunctionTok{sobolSalt}\NormalTok{(}\AttributeTok{model =}\NormalTok{ ishigami.fun, }\AttributeTok{X1 =}\NormalTok{ mat[}\DecValTok{1}\SpecialCharTok{:}\NormalTok{N, ], }
                               \AttributeTok{X2 =}\NormalTok{ mat[(N}\SpecialCharTok{+}\DecValTok{1}\NormalTok{)}\SpecialCharTok{:}\NormalTok{(}\DecValTok{2}\SpecialCharTok{*}\NormalTok{N), ], }\AttributeTok{nboot =}\NormalTok{ R)}
\FunctionTok{print}\NormalTok{(}\FunctionTok{round}\NormalTok{(sensitivity\_ind}\SpecialCharTok{$}\NormalTok{S, }\DecValTok{3}\NormalTok{))}
\end{Highlighting}
\end{Shaded}

\begin{verbatim}
##    original bias std. error min. c.i. max. c.i.
## X1    0.313    0      0.012     0.289     0.336
## X2    0.442    0      0.011     0.421     0.465
## X3   -0.002    0      0.019    -0.040     0.035
\end{verbatim}

\begin{Shaded}
\begin{Highlighting}[]
\FunctionTok{print}\NormalTok{(}\FunctionTok{round}\NormalTok{(sensitivity\_ind}\SpecialCharTok{$}\NormalTok{T, }\DecValTok{3}\NormalTok{))}
\end{Highlighting}
\end{Shaded}

\begin{verbatim}
##    original bias std. error min. c.i. max. c.i.
## X1    0.552    0      0.017     0.515     0.584
## X2    0.445    0      0.011     0.423     0.465
## X3    0.243    0      0.006     0.232     0.255
\end{verbatim}

\hypertarget{Shapley-effect}{%
\section{SA in the case of correlated inputs}\label{Shapley-effect}}

In many applications such as epidemiology it happens that the input
parameters are correlated. In this situation, the Sobol' indices are not
reliable sensitivity measures because they may over/underestimate the
contribution of each input parameter on the output variability
(\protect\hyperlink{ref-owen2014}{Owen, 2014}) since the variance decomposition in \eqref{eq:var-decompose}
relies on the assumption that the factors are independent. More
precisely, Equation \eqref{eq:first-total-relation} may not hold in the
case of dependent inputs, e.g.~\(\sum_{i = 1}^{d}S_i > 1\) or
\(\sum_{i = 1}^{d}S_{T_i} < 1\) (\protect\hyperlink{ref-song2016}{Song et al., 2016}). For example, for the model
below \begin{equation}
Y = X_1 + X_2,~ \begin{pmatrix}
X_1 \\
X_2 
\end{pmatrix} \sim \mathcal{N} \left(\begin{pmatrix}
\mu_1 \\
\mu_2, \end{pmatrix}, \begin{pmatrix}
\sigma^2_1 & \rho\sigma_1\sigma_2 \\
\rho\sigma_1\sigma_2 & \sigma^2_2 \end{pmatrix}
\right), ~ \sigma_1 = \sigma_2 > 0, 
\label{eq:2D-Shaplry-example}
\end{equation} we have \begin{align}
\mathbb{E}\lbrack Y\mid X_1\rbrack &= X_1 + \mu_2 + \frac{\rho\sigma_2}{\sigma_1} \left(X_1 - \mu_1\right) \implies
\mathbb{V}\left(\mathbb{E}\lbrack Y\mid X_1\rbrack \right) = \sigma^2_1 \left(1 + \rho \right)^2 \\
\mathbb{V}(Y) &= \sigma^2_1 + 2\rho\sigma_1\sigma_2 + \sigma^2_2 = 2\sigma^2_1(1 + \rho) \\
S_1 &= S_2 = \frac{\sigma^2_1(1 + \rho)^2}{2\sigma^2_1(1 + \rho)} = \frac{1 + \rho}{2} . 
\end{align} As can be seen, if the correlation coefficient \(\rho\) is
positive, the sum of \(S_1\) and \(S_2\) is greater than one. To overcome
this problem, Owen (\protect\hyperlink{ref-owen2014}{Owen, 2014}) proposed to use the Shapley value/effect
which is a concept from cooperative game theory (\protect\hyperlink{ref-shapley1953}{Shapley, 1953}). It
offers a unique and fair solution to the problem of distributing a
game's total gain/payoff among the players according to their relative
contribution. Using the Shapley effect, a single positive sensitivity
index (instead of two indices) is obtained for each input making the
interpretation easier than the Sobol' method. Besides, the sum of the
Shapley values is equal to one even if there is a strong correlation
among the factors. In the following, we first introduce briefly the game
theory idea behind the Shapley value and then focus on its application
in SA of simulators with dependent inputs.

A cooperative game is characterised by a set of players
\(\mathcal{D} = \lbrace 1, \ldots, d\rbrace\) called the ``grand coalition''
and a characteristic function \(\nu : 2^{\mathcal{D}}\mapsto \mathbb{R}\).
The coalition \(\mathcal{D}\) can be viewed as the set of all input
factors in the SA paradigm. The characteristic function maps subsets of
players \(I \subseteq \mathcal{D}\) to a real number \(\nu(I)\) which
reflects the payoff that the members of the coalition \(I\) can achieve by
cooperation. Notice that the empty set (\(I = \emptyset\)) is also a
coalition with a zero payoff (\(\nu(\emptyset) = 0\)). Now let
\(\mathcal{J} \subseteq \mathcal{D}\setminus \{i\}\) represent a coalition
of \(\lvert \mathcal{J} \rvert\) players not containing the player \(i\).
The marginal contribution of the player \(i\) with respect to \(\nu(\cdot)\)
is given by
\(\nu\left(\mathcal{J}\cup \{i\}\right) - \nu\left(\mathcal{J}\right)\)
which indicates the incremental value for including the player \(i\) in
the coalition \(\mathcal{J}\). The Shapley value is then expressed by
\begin{equation}
\phi_i = \sum_{\mathcal{J}} \frac{\lvert \mathcal{J} \rvert! \left(d - \lvert \mathcal{J} \rvert - 1 \right)!}{d !} \lbrack \nu\left(\mathcal{J}\cup \{i\}\right) - \nu\left(\mathcal{J}\right)\rbrack , ~ i = 1, \ldots, d ,
\label{eq:shap-val}
\end{equation} that is a weighted average of the marginal contribution
of the player \(i\) over all possible coalitions, including
\(\mathcal{J} = \emptyset\). The Shapley value can be computed in a
different manner, which is based on the permutations of \(\mathcal{D}\)
\begin{equation}
\phi_i = \frac{1}{d !}\sum_{\pi\in\Pi} \lbrack \nu\left(P_i(\pi)\cup \{i\}\right) - \nu\left(P_i(\pi)\right)\rbrack , ~ i = 1, \ldots, d ,
\label{eq:shap-val-permut}
\end{equation} where \(\Pi\) is the set of all \(d !\) permutations of
players and \(P_i(\pi)\) is the set of players that precedes player \(i\) in
the permutation \(\pi\in\Pi\).

The main disadvantage of the Shapley value is its enormous computational
burden, especially when \(d\) is large. The reason is that computing the
Shapley effect requires all possible subsets of players. To overcome
this complexity, various approximation methods have been developed. For
example, Castro et al. (\protect\hyperlink{ref-castro2009}{Castro et al., 2009}) proposed the following expression
relying on Equation \eqref{eq:shap-val-permut} to approximate the Shapley
value \begin{equation}
\hat{\phi}_i = \frac{1}{M} \sum_{m = 1}^{M} \lbrack \nu\left(P_i(\pi_m)\cup \{i\}\right) - \nu\left(P_i(\pi_m)\right)\rbrack , ~ i = 1, \ldots, d .
\label{eq:approx-shap-val-permut}
\end{equation} In the above equation, \(\pi_1, \ldots, \pi_M\) are \(M\)
random permutations in \(\Pi\).

Now we describe how the Shapley values can be used for the sensitivity
study of computer codes specially if the inputs are dependent. In this
framework, the model factors are deemed as players of a game with the
total payoff \(\mathbb{V}(Y)\) (or one in the normalized case). Also, for
a set of inputs \(I \subseteq D\), \(\nu\left(I\right)\) returns the output
uncertainty caused by the uncertainty of those inputs. Thus, the Shapley
value is a variance-based method and serves as a global sensitivity
metric.

Two possible choices of the characteristic function are (\protect\hyperlink{ref-iooss2019}{Iooss \& Prieur, 2019})
\begin{align}
\label{eq:characteristic-sa1}
&\nu_1\left(I\right) = \frac{\mathbb{V}_{\mathbf{X}_I}\left(\mathbb{E}\lbrack Y\mid\mathbf{X}_I \rbrack\right)}{\mathbb{V}(Y)} ,  \\
&\nu_2\left(I\right) = \frac{\mathbb{E}_{\mathbf{X}_{\sim I}}\left(\mathbb{V}\lbrack Y\mid\mathbf{X}_{\sim I} \rbrack\right)}{\mathbb{V}(Y)} , 
\label{eq:characteristic-sa2}
\end{align} where the former is interpreted as the first order and the
latter as the total order effect in the Sobol' formulation. Although it
is proved that both characteristic functions yield the same Shapley
value (\protect\hyperlink{ref-song2016}{Song et al., 2016}), the MC estimation to \(\nu_2\) is always unbiased
making it a more popular choice. In contrast, the estimator of \(\nu_1\)
can be badly biased if the number of MC samples used to evaluate the
conditional expectation (i.e.
\(\mathbb{E}\lbrack Y\mid\mathbf{X}_I \rbrack\)) is small (\protect\hyperlink{ref-radaiideh2019}{Radaideh et al., 2019}; \protect\hyperlink{ref-sun2011}{Sun et al., 2011}). As recommended in (\protect\hyperlink{ref-song2016}{Song et al., 2016}; \protect\hyperlink{ref-sun2011}{Sun et al., 2011}), the estimation of
\(\nu_2\) is performed via a two-level MC: an inner loop
for the conditional variance and an outer loop for the
expectation estimation. According to the theoretical analysis in
(\protect\hyperlink{ref-song2016}{Song et al., 2016}), it is suggested that a suitable size for the inner and outer loops is one and three, respectively. This algorithm is implemented in the package \textbf{sensitivity} with the
characteristic function \(\nu_2\). More recent algorithms (see e.g., (\protect\hyperlink{ref-broto2020}{Broto et al., 2020}, \protect\hyperlink{ref-broto2022}{2022}; \protect\hyperlink{ref-daveiga2021}{Da Veiga et al., 2021})) for the estimation of the Shapley effects are also included in the \textbf{sensitivity} package. They are implemented in the functions \texttt{shapleySubsetMc()}, \texttt{shapleysobol\_knn()}, and \texttt{sobolshap\_knn()}.

\hypertarget{stochastic-model}{%
\section{Case of stochastic models}\label{stochastic-model}}

Stochastic simulators such as agent-based models are ubiquitous in the
social and biological sciences (\protect\hyperlink{ref-binois2018}{Binois et al., 2018}). In stochastic models,
contrary to deterministic ones, different observations are attained at
an identical input due to the inclusion of a random number seed in their
code (\protect\hyperlink{ref-ohagan2006}{O'Hagan, 2006}). To characterise the response distribution at a
specific input, we need to run the code with the same input repeatedly.
Hence, applying SA to stochastic
simulators requires a larger number of model
evaluations than the case of deterministic codes. If the stochastic model is computationally expensive, conducting SA becomes impossible. To address this issue, one can replace the simulator with a
cheap-to-evaluate surrogate model and perform SA on it. In this framework, there are different classes of surrogate models; see e.g., (\protect\hyperlink{ref-sudret2008}{Sudret, 2008}; \protect\hyperlink{ref-zhu2021}{Zhu \& Sudret, 2021}). Here, we only consider Gaussian processes (GP) emulators
(\protect\hyperlink{ref-GPML}{Rasmussen \& Williams, 2005}). GPs have become the gold standard surrogate model in the
field of the design and analysis of computer experiments due to their
statistical properties (\protect\hyperlink{ref-santner2003}{Santner et al., 2003}). For example, the GP prediction is equipped with an estimation
of uncertainty that reflects the accuracy of the prediction. Some applications of GPs in modelling computer experiments
can be found in (\protect\hyperlink{ref-beck2016}{Beck \& Guillas, 2016}; \protect\hyperlink{ref-mohammadi2019}{Mohammadi et al., 2019}; \protect\hyperlink{ref-vernon2018}{Vernon et al., 2018}). The statistical background of GPs is presented below.

\hypertarget{GP-emulator}{%
\subsection{Gaussian process emulators}\label{GP-emulator}}

We consider
the output of a stochastic model to be of the following form
\begin{equation}
    y(\mathbf{x}) = f(\mathbf{x}) + \varepsilon , ~ \varepsilon \sim \mathcal{N}\left(0, \tau(\mathbf{x})\right) .
    \label{eq:stochastic-model}
\end{equation} The above expression represents the general
\emph{heteroscedastic} case as the noise variance \(\tau(\mathbf{x})\)
(also referred to as the \emph{nugget} (\protect\hyperlink{ref-binois2018}{Binois et al., 2018})) changes across the
input space. If the
noise variance is constant, the model is called \emph{homoscedastic}. In the GP paradigm, the prior belief about the form of
\(f\) is modelled via the stochastic process \begin{equation}
    Z_{\mathbf{x}} = \mu(\mathbf{x}) + \eta_{\mathbf{x}} , 
    \label{eq:stat-model}
\end{equation} where \(\mu(\mathbf{x})\) is the \emph{trend function} and
\(\eta_{\mathbf{x}}\) is a centred (or zero mean) GP. Without loss of
generality, we assume that the trend function is a constant denoted by
\(\mu_0\). The covariance structure of \(\eta_{\mathbf{x}}\) is determined
by its positive definite covariance function/kernel \(c(\cdot, \cdot)\)
defined as \begin{equation}
    c: \mathbb{R}^d \times \mathbb{R}^d \mapsto \mathbb{R}, ~ c\left(\mathbf{x}, \mathbf{x}^\prime \right) = \mathbb{C}\text{ov}\left(\eta_{\mathbf{x}}, \eta_{\mathbf{x}^\prime} \right) . 
    \label{eq:kernel}
\end{equation} Although there are many options available for the choice
of the covariance function, the Matern or squared exponential kernels
(\protect\hyperlink{ref-GPML}{Rasmussen \& Williams, 2005}) are typically adopted in the computer experiments literature.
Traditionally, a parameterized family of \(c\) is specified and its
parameters are estimated from the data by e.g., maximum likelihood
(\protect\hyperlink{ref-roustant2012}{Roustant et al., 2012}).

Now let
\(\bm{\mathcal{X}}_n = \left(\mathbf{x}^{(1)}, \ldots, \mathbf{x}^{(n)}\right)^\top\)
be \(n\) locations (called the design points) in the input space with the
corresponding noisy output observations
\(\bm{\mathcal{Y}}_n = \left(y\left(\mathbf{x}^{(1)}\right), \ldots, y\left(\mathbf{x}^{(n)}\right)\right)^\top\).
Often, the elements of \(\bm{\mathcal{X}}_n\) are selected according to a
space-filling design. Given that all parameters in Equation
\eqref{eq:stat-model} are known, the predictive distribution at any site
\(\mathbf{x}^\ast\) is driven by the posterior distribution
\(Z_{\mathbf{x}^\ast} \mid \bm{\mathcal{Y}}_n\) which is Gaussian
charactrised by \begin{align}
    \label{eq:GP-mean}
    &\mu(\mathbf{x}^\ast) = \mathbb{E}\lbrack Z_{\mathbf{x}^\ast} \mid \bm{\mathcal{Y}}_n \rbrack = \mu_0 + \mathbf{c}^\top \mathbf{C}^{-1} \left(\bm{\mathcal{X}}_n - \mu_0\mathbf{1}\right) , \\
    &\sigma^2(\mathbf{x}^\ast) = \mathbb{V}\lbrack Z_{\mathbf{x}^\ast} \mid \bm{\mathcal{Y}}_n \rbrack = c(\mathbf{x}^\ast, \mathbf{x}^\ast) + \tau(\mathbf{x}^\ast) - \mathbf{c}^\top \mathbf{C}^{-1} \mathbf{c} .
    \label{eq:GP-var}
\end{align} Here,
\(\mathbf{c} = \left(c\left(\mathbf{x}^\ast, \mathbf{x}^{(1)}\right), \ldots, c\left(\mathbf{x}^\ast, \mathbf{x}^{(n)}\right)\right)^\top\)
and \(\mathbf{C}\) is an \(n\times n\) covariance matrix whose elements are
\(\mathbf{C}_{kl} = c\left(\mathbf{x}^{(k)}, \mathbf{x}^{(l)}\right) + \delta_{lk}\tau \left(\mathbf{x}^{(k)}\right)\)
where \(\delta_{lk}\) is the Kronecker delta function, for
\(1 \le l,k \le n\). The GP predictive mean and variance expressions for
deterministic codes are analogous to Equations \eqref{eq:GP-mean} and
\eqref{eq:GP-var} except that the noise variance term is discarded.

\begin{figure}

{\centering \includegraphics{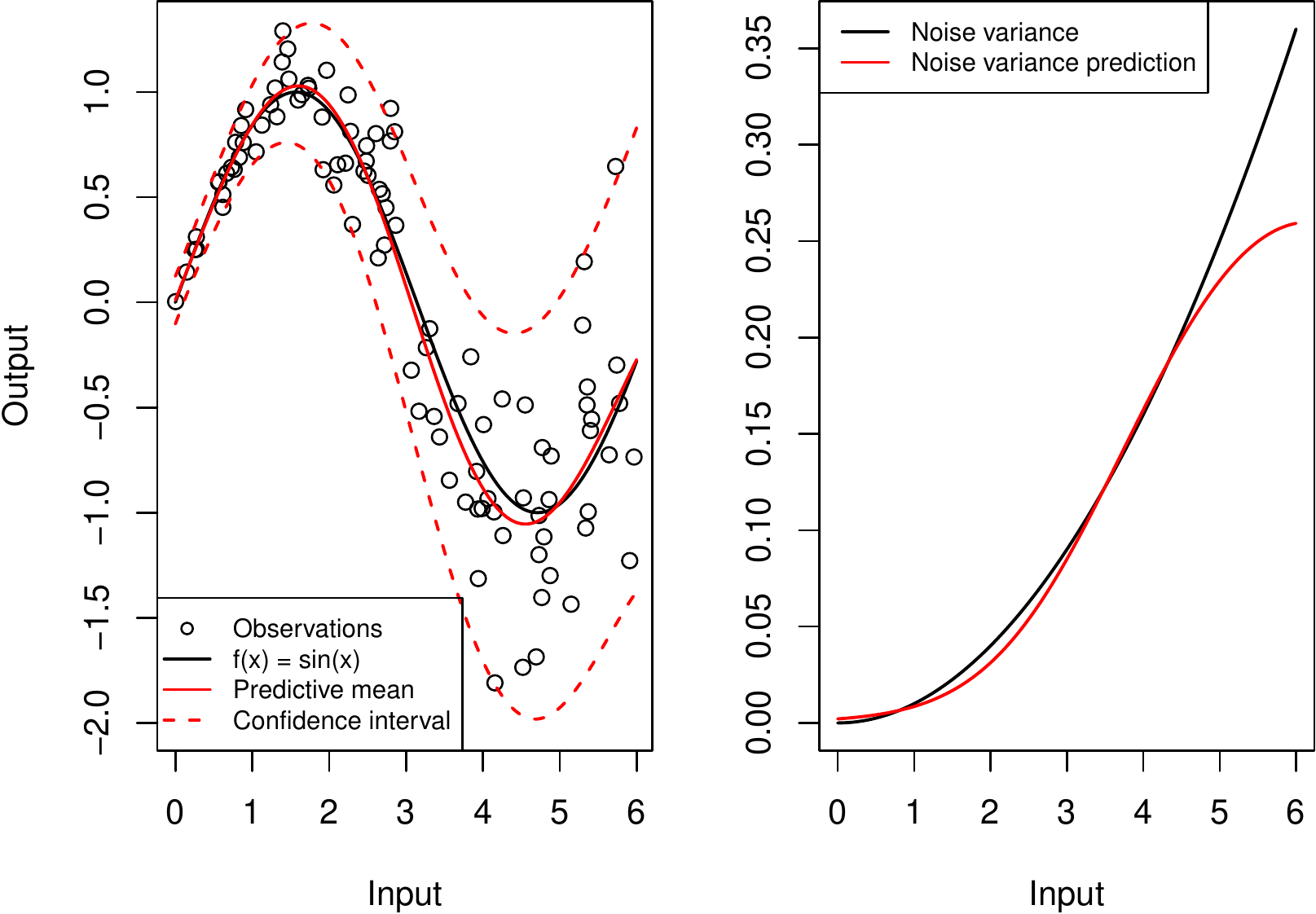} 

}

\caption{Left: a heteroscedastic stochastic model defined by: $y(x) = \sin(x) + \varepsilon, \varepsilon \sim \mathcal{N}\left(0, \tau(x) = 0.01x^2\right), x\in[0, 6]$. There are 100 observations (black circles) taken from $y(x)$. The GP emulator (red) is constructed using the function \texttt{mleHetGP} implemented in the package \textbf{hetGP}. Right: the noise variance (black) and its prediction, $\hat{\tau}(x)$ (red).}\label{fig:hetGP}
\end{figure}

In practice, the true value of \(\tau(\mathbf{x})\) is unknown. The noise
variance can be estimated at the design points by repeatedly running the
simulator there and computing the sample variances. This method is called the
\emph{stochastic kriging} (\protect\hyperlink{ref-ankenman2010}{Ankenman et al., 2010}); it only works if there are enough
replicated observations at \(\mathbf{x}^{(k)}, k = 1, \ldots, n\). Yet, it
is not possible to estimate \(\tau(\mathbf{x}^\ast)\) in the GP predictive
variance (Equation \eqref{eq:GP-var}) as no observations are available at
\(\mathbf{x}^\ast\). To overcome this problem, Binois et al. (\protect\hyperlink{ref-binois2018}{Binois et al., 2018})
proposed a computationally efficient method such that a joint GP model is used
to fit the mean response and noise variance. In this approach, the
noise variances at the design points,
\(\left(\tau\left(\mathbf{x}^{(1)}\right), \ldots, \tau\left(\mathbf{x}^{(n)}\right)\right)\),
are treated as latent variables that can be learnt together with the
kernel parameters through a joint likelihood. An implementation of this
method is available in the \textbf{R} package \textbf{hetGP} (\protect\hyperlink{ref-binois2021}{Binois \& Gramacy, 2021}). Figure
\ref{fig:hetGP} visualises a heteroscedastic example in which the true
function and the noise variance are \(f(x) = \sin(x)\) and
\(\tau(x) = 0.01x^2\), respectively. On the left picture of Figure
\ref{fig:hetGP}, the emulator (red) is built based on 100 noisy
observations (black circles): predictive mean (solid) and confidence
interval (dashed). The red line on the right panel shows the noise
variance prediction, \(\hat{\tau}(x)\).

\begin{Shaded}
\begin{Highlighting}[]
\FunctionTok{library}\NormalTok{(hetGP)}
\FunctionTok{set.seed}\NormalTok{(}\DecValTok{123}\NormalTok{)}
\NormalTok{tau }\OtherTok{\textless{}{-}} \ControlFlowTok{function}\NormalTok{(xx) }\FloatTok{0.01}\SpecialCharTok{*}\NormalTok{xx}\SpecialCharTok{\^{}}\DecValTok{2}  \CommentTok{\# Noise variance function}
\NormalTok{fun }\OtherTok{\textless{}{-}} \ControlFlowTok{function}\NormalTok{(xx) \{}
\NormalTok{  yy }\OtherTok{\textless{}{-}} \FunctionTok{sin}\NormalTok{(xx) }\SpecialCharTok{+} \FunctionTok{rnorm}\NormalTok{(}\DecValTok{1}\NormalTok{, }\DecValTok{0}\NormalTok{, }\FunctionTok{sqrt}\NormalTok{(}\FunctionTok{tau}\NormalTok{(xx)))}
  \FunctionTok{return}\NormalTok{(yy)}
\NormalTok{\}}
\NormalTok{X\_n }\OtherTok{\textless{}{-}} \FunctionTok{as.matrix}\NormalTok{(}\FunctionTok{runif}\NormalTok{(}\DecValTok{100}\NormalTok{, }\DecValTok{0}\NormalTok{, }\DecValTok{6}\NormalTok{))  }\CommentTok{\# Design points}
\NormalTok{Y\_n }\OtherTok{\textless{}{-}} \FunctionTok{apply}\NormalTok{(X\_n, }\DecValTok{1}\NormalTok{, fun) }\CommentTok{\# Outputs at X\_n  }
\NormalTok{emulator }\OtherTok{\textless{}{-}} \FunctionTok{mleHetGP}\NormalTok{(}\AttributeTok{X =}\NormalTok{ X\_n, }\AttributeTok{Z =}\NormalTok{ Y\_n, }\AttributeTok{lower =} \FloatTok{0.1}\NormalTok{, }\AttributeTok{upper =} \DecValTok{20}\NormalTok{,}
                     \AttributeTok{maxit =} \DecValTok{1000}\NormalTok{, }\AttributeTok{covtype =} \StringTok{"Gaussian"}\NormalTok{)}
\NormalTok{x }\OtherTok{\textless{}{-}} \FunctionTok{as.matrix}\NormalTok{(}\FunctionTok{seq}\NormalTok{(}\DecValTok{0}\NormalTok{, }\DecValTok{6}\NormalTok{, }\FloatTok{0.01}\NormalTok{))}
\NormalTok{P }\OtherTok{\textless{}{-}} \FunctionTok{predict}\NormalTok{(}\AttributeTok{object =}\NormalTok{ emulator, }\AttributeTok{x =}\NormalTok{ x)}
\NormalTok{CI }\OtherTok{\textless{}{-}} \DecValTok{2}\SpecialCharTok{*}\FunctionTok{sqrt}\NormalTok{(P}\SpecialCharTok{$}\NormalTok{sd2 }\SpecialCharTok{+}\NormalTok{ P}\SpecialCharTok{$}\NormalTok{nugs)  }\CommentTok{\# Confidence interval }
\CommentTok{\#P$sd2: predictive variance, P$nugs: noise variance prediction}
\FunctionTok{par}\NormalTok{(}\AttributeTok{mfcol =} \FunctionTok{c}\NormalTok{(}\DecValTok{1}\NormalTok{, }\DecValTok{2}\NormalTok{))}
\FunctionTok{par}\NormalTok{(}\AttributeTok{mar =} \FunctionTok{c}\NormalTok{(}\DecValTok{4}\NormalTok{, }\DecValTok{4}\NormalTok{, .}\DecValTok{1}\NormalTok{, .}\DecValTok{1}\NormalTok{))}
\FunctionTok{plot}\NormalTok{(X\_n, Y\_n, }\AttributeTok{col =} \StringTok{"black"}\NormalTok{, }\AttributeTok{xlab =} \StringTok{"Input"}\NormalTok{, }\AttributeTok{ylab =} \StringTok{"Output"}\NormalTok{, }\AttributeTok{ylim =} \FunctionTok{c}\NormalTok{(}\SpecialCharTok{{-}}\DecValTok{2}\NormalTok{, }\FloatTok{1.3}\NormalTok{))}
\FunctionTok{lines}\NormalTok{(x, }\FunctionTok{sin}\NormalTok{(x), }\AttributeTok{type =} \StringTok{"l"}\NormalTok{, }\AttributeTok{lwd =} \FloatTok{1.5}\NormalTok{)}
\FunctionTok{lines}\NormalTok{(x, P}\SpecialCharTok{$}\NormalTok{mean, }\AttributeTok{col =} \StringTok{"red"}\NormalTok{, }\AttributeTok{lwd =} \FloatTok{1.5}\NormalTok{)  }\CommentTok{\# P$mean: predictive mean}
\FunctionTok{lines}\NormalTok{(x, P}\SpecialCharTok{$}\NormalTok{mean }\SpecialCharTok{+}\NormalTok{ CI, }\AttributeTok{col =} \StringTok{"red"}\NormalTok{, }\AttributeTok{lty =} \DecValTok{2}\NormalTok{, }\AttributeTok{lwd =} \FloatTok{1.5}\NormalTok{)}
\FunctionTok{lines}\NormalTok{(x, P}\SpecialCharTok{$}\NormalTok{mean }\SpecialCharTok{{-}}\NormalTok{ CI, }\AttributeTok{col =} \StringTok{"red"}\NormalTok{, }\AttributeTok{lty =} \DecValTok{2}\NormalTok{, }\AttributeTok{lwd =} \FloatTok{1.5}\NormalTok{)}
\FunctionTok{legend}\NormalTok{(}\StringTok{"bottomleft"}\NormalTok{, }\AttributeTok{lwd=}\FunctionTok{c}\NormalTok{(}\DecValTok{1}\NormalTok{,}\FloatTok{1.5}\NormalTok{,}\FloatTok{1.5}\NormalTok{,}\FloatTok{1.5}\NormalTok{), }\AttributeTok{lty=}\FunctionTok{c}\NormalTok{(}\ConstantTok{NA}\NormalTok{,}\DecValTok{1}\NormalTok{,}\DecValTok{1}\NormalTok{,}\DecValTok{2}\NormalTok{), }\AttributeTok{pch=}\FunctionTok{c}\NormalTok{(}\DecValTok{1}\NormalTok{,}\ConstantTok{NA}\NormalTok{,}\ConstantTok{NA}\NormalTok{,}\ConstantTok{NA}\NormalTok{), }
       \AttributeTok{legend=}\FunctionTok{c}\NormalTok{(}\StringTok{"Observations"}\NormalTok{,}\StringTok{"f(x) = sin(x)"}\NormalTok{,}\StringTok{"Predictive mean"}\NormalTok{,}\StringTok{"Confidence interval"}\NormalTok{), }
       \AttributeTok{col =} \FunctionTok{c}\NormalTok{(}\StringTok{"black"}\NormalTok{, }\StringTok{"black"}\NormalTok{, }\StringTok{"red"}\NormalTok{, }\StringTok{"red"}\NormalTok{, }\StringTok{"red"}\NormalTok{), }\AttributeTok{cex =} \FloatTok{0.75}\NormalTok{)}
\FunctionTok{plot}\NormalTok{(x, }\FunctionTok{tau}\NormalTok{(x), }\AttributeTok{type =} \StringTok{"l"}\NormalTok{, }\AttributeTok{lwd =} \FloatTok{1.5}\NormalTok{, }\AttributeTok{xlab =} \StringTok{"Input"}\NormalTok{, }\AttributeTok{ylab =} \StringTok{""}\NormalTok{)}
\FunctionTok{lines}\NormalTok{(x, P}\SpecialCharTok{$}\NormalTok{nugs, }\AttributeTok{col =} \StringTok{"red"}\NormalTok{, }\AttributeTok{lwd =} \FloatTok{1.5}\NormalTok{)}
\FunctionTok{legend}\NormalTok{(}\StringTok{"topleft"}\NormalTok{, }\AttributeTok{legend =} \FunctionTok{c}\NormalTok{(}\StringTok{"Noise variance"}\NormalTok{, }\StringTok{"Noise variance prediction"}\NormalTok{), }
       \AttributeTok{col =} \FunctionTok{c}\NormalTok{(}\StringTok{"black"}\NormalTok{, }\StringTok{"red"}\NormalTok{), }\AttributeTok{lwd =} \FloatTok{1.5}\NormalTok{, }\AttributeTok{lty =} \DecValTok{1}\NormalTok{, }\AttributeTok{cex =} \FloatTok{0.75}\NormalTok{)}
\end{Highlighting}
\end{Shaded}

\hypertarget{SA-stochastic-model}{%
\subsection{SA of stochastic models}\label{SA-stochastic-model}}

One way to conduct SA for stochastic simulators is to extend the set of
the input parameters \(\mathbf{X}\) by an extra variable \(X_\varepsilon\)
which denotes the uncontrollable parameter governed by the simulator
itself (\protect\hyperlink{ref-iooss2009}{Iooss \& Ribatet, 2009}). The new variable \(X_\varepsilon\) is called the
``seed variable'' and is independent of the other inputs. Accordingly, one
can write the output variable as \begin{equation}
Y = f\left(\mathbf{X}, X_\varepsilon \right) ,
\end{equation} meaning that the response variability consists of
intrinsic randomness caused by the seed variable and uncertainty in the
inputs. In this framework, the mean (\(Y_m\)) and variance (\(Y_v\))
function of stochastic simulators are given by \begin{align}
\label{eq:simulator-mean}
&Y_m\left(\mathbf{X}\right) = \mathbb{E}_{X_\varepsilon} \lbrack Y \mid \mathbf{X}\rbrack, \\
&Y_v\left(\mathbf{X}\right) = \mathbb{V}_{X_\varepsilon}\left(Y \mid \mathbf{X}\right) = \mathbb{E}_{X_\varepsilon} \lbrack \left(Y - Y_m\left(\mathbf{X}\right)\right)^2 \mid \mathbf{X}\rbrack .
\label{eq:simulator-var}
\end{align}

It is shown that the first order effect of each \(X_i\) (\(S_i\)) and the
total order index of \(X_\varepsilon\) (\(S_{T_\varepsilon}\)) can be
expressed in terms of \(Y_m\) and \(Y_v\) (\protect\hyperlink{ref-iooss2009}{Iooss \& Ribatet, 2009}; \protect\hyperlink{ref-marrel2012}{Marrel et al., 2012}). Thanks
to the law of total expectation, we have \begin{equation}
\mathbb{E}_{\mathbf{X}_{\sim i}}\lbrack Y \mid X_i \rbrack = \mathbb{E}_{\mathbf{X}_{\sim i}}\lbrack \mathbb{E}_{X_\varepsilon}\lbrack Y \mid \mathbf{X}\rbrack \mid X_i \rbrack = \mathbb{E}_{\mathbf{X}_{\sim i}}\lbrack Y_m\left(\mathbf{X}\right) \mid X_i \rbrack .
\end{equation} Consequently, the first order effect of \(X_i\) can be
rewritten as \begin{equation}
S_i = \frac{\mathbb{V}_{X_i} \left(\mathbb{E}_{\mathbf{X}_{\sim i}} \lbrack Y \mid X_i\rbrack \right)}{\mathbb{V}(Y)} = \frac{\mathbb{V}_{X_i} \left(\mathbb{E}_{\mathbf{X}_{\sim i}}\lbrack Y_m\left(\mathbf{X}\right) \mid X_i\rbrack \right)}{\mathbb{V}(Y)} ,
\end{equation} which relies on the mean response function. The total
order effect of \(X_\varepsilon\) takes the following form
\begin{equation}
S_{T_\varepsilon} = \frac{\mathbb{E}_{\mathbf{X}} \left[\mathbb{V}_{X_\varepsilon} \left(Y \mid \mathbf{X}\right) \right]}{\mathbb{V}(Y)} = \frac{\mathbb{E}_{\mathbf{X}} \left[Y_v\left(\mathbf{X}\right) \right]}{\mathbb{V}(Y)} , 
\label{eq:total-effect-stochastic}
\end{equation} referring to the total sensitivity index defined in
Equation \eqref{eq:total-effect}. Moreover, the output variance can be
expressed as a function of \(Y_m\) and \(Y_v\) thanks to the law of total
variance: \begin{equation}
\mathbb{V}\left(Y\right) = \mathbb{V}_{\mathbf{X}}\left(\mathbb{E}_{X_\varepsilon}\lbrack Y \mid \mathbf{X}\rbrack\right) + \mathbb{E}_{\mathbf{X}}\lbrack \mathbb{V}_{X_\varepsilon}\left(Y \mid \mathbf{X}\right)\rbrack = \mathbb{V}_{\mathbf{X}}\left(Y_m\left(\mathbf{X}\right) \right) + \mathbb{E}_{\mathbf{X}}\lbrack Y_v\left(\mathbf{X}\right) \rbrack .
\end{equation}

The advantage of expressing the sensitivity indices in terms of \(Y_m\)
and \(Y_v\) is that they can be approximated by GP emulators. This leads
to a significant reduction in the computational cost of conducting SA
for stochastic simulators. While \(Y_m\) is approximated by the GP
predictive mean (Equation \eqref{eq:GP-mean}), the prediction of \(Y_v\) in
the heteroscedastic case needs more careful attention. To tackle this
problem, (\protect\hyperlink{ref-marrel2012}{Marrel et al., 2012}) suggested a joint surrogate modelling approach
which requires constructing several GP emulators. However, one can use
the noise variance prediction offered by the package \textbf{hetGP}
(\protect\hyperlink{ref-binois2021}{Binois \& Gramacy, 2021}) to approximate \(Y_v\) as explained in Section
\ref{GP-emulator}. Finally, it is worth mentioning recent alternatives for SA of stochastic codes such as SA in Wasserstein spaces (\protect\hyperlink{ref-fort2021}{Fort et al., 2021}), or kernel-based SA (\protect\hyperlink{ref-daveiga2021a}{Da Veiga, 2021}).

\hypertarget{conclusion}{%
\section{Conclusion}\label{conclusion}}

In this report, we investigated various aspects of sensitivity analysis
of numerical models that one can encounter in real-world applications.
This includes the Sobol' indices, SA of stochastic simulators and those
with dependent inputs. The latter is tackled via the Shapley effect
since the Sobol' indices are not reliable measures when the inputs are
correlated. The Shapley effect is a concept in cooperative game theory.
In the case of stochastic simulators, we first employed a GP to emulate
the model and then applied SA on the emulator. GPs are commonplace
surrogate models in the field of computer experiments to alleviate the
computational burden. The analysis is carried out (mainly) with \textbf{R}
packages \textbf{sensitivity} and \textbf{sensobol}. We provided several
illustrative examples that help the user to learn the packages easily.
All the results are reproducible making the report important from a
practical point of view.

\hypertarget{acknowledgements}{%
\section*{Acknowledgements}\label{acknowledgements}}
\addcontentsline{toc}{section}{Acknowledgements}

The authors (HM and PC) would like to thank the Alan Turing Institute for funding
this work.

\hypertarget{references}{%
\section*{References}\label{references}}
\addcontentsline{toc}{section}{References}

\hypertarget{refs}{}
\begin{CSLReferences}{1}{0}
\leavevmode\vadjust pre{\hypertarget{ref-ankenman2010}{}}%
Ankenman, B., Nelson, B. L., \& Staum, J. (2010). Stochastic kriging for simulation metamodeling. \emph{Operations Research}, \emph{58}(2), 371--382.

\leavevmode\vadjust pre{\hypertarget{ref-beck2016}{}}%
Beck, J., \& Guillas, S. (2016). Sequential design with mutual information for computer experiments {(MICE)}: Emulation of a tsunami model. \emph{SIAM/ASA Journal on Uncertainty Quantification}, \emph{4}(1), 739--766. \url{https://doi.org/10.1137/140989613}

\leavevmode\vadjust pre{\hypertarget{ref-binois2021}{}}%
Binois, M., \& Gramacy, R. B. (2021). {hetGP: Heteroskedastic Gaussian process modeling and sequential design in R}. \emph{Journal of Statistical Software}, \emph{98}(13), 1--44. \url{https://doi.org/10.18637/jss.v098.i13}

\leavevmode\vadjust pre{\hypertarget{ref-binois2018}{}}%
Binois, M., Gramacy, R. B., \& Ludkovski, M. (2018). Practical heteroscedastic {G}aussian process modeling for large simulation experiments. \emph{Journal of Computational and Graphical Statistics}, \emph{27}(4), 808--821. \url{https://doi.org/10.1080/10618600.2018.1458625}

\leavevmode\vadjust pre{\hypertarget{ref-borgonovo2016}{}}%
Borgonovo, E., \& Plischke, E. (2016). Sensitivity analysis: A review of recent advances. \emph{European Journal of Operational Research}, \emph{248}(3), 869--887. https://doi.org/\url{https://doi.org/10.1016/j.ejor.2015.06.032}

\leavevmode\vadjust pre{\hypertarget{ref-broto2022}{}}%
Broto, B., Bachoc, F., Clouvel, L., \& Martinez, J.-M. (2022). Block-diagonal covariance estimation and application to the {S}hapley effects in sensitivity analysis. \emph{SIAM/ASA Journal on Uncertainty Quantification}, \emph{10}(1), 379--403. \url{https://doi.org/10.1137/20M1358839}

\leavevmode\vadjust pre{\hypertarget{ref-broto2020}{}}%
Broto, B., Bachoc, F., \& Depecker, M. (2020). Variance reduction for estimation of {S}hapley effects and adaptation to unknown input distribution. \emph{SIAM/ASA Journal on Uncertainty Quantification}, \emph{8}(2), 693--716. \url{https://doi.org/10.1137/18M1234631}

\leavevmode\vadjust pre{\hypertarget{ref-burnaev2017}{}}%
Burnaev, E., Panin, I., \& Sudret, B. (2017). Efficient design of experiments for sensitivity analysis based on polynomial chaos expansions. \emph{Annals of Mathematics and Artificial Intelligence}, \emph{81}(1), 187--207. \url{https://doi.org/10.1007/s10472-017-9542-1}

\leavevmode\vadjust pre{\hypertarget{ref-castro2009}{}}%
Castro, J., Gómez, D., \& Tejada, J. (2009). {Polynomial calculation of the Shapley value based on sampling}. \emph{Computers \& Operations Research}, \emph{36}(5), 1726--1730. https://doi.org/\url{https://doi.org/10.1016/j.cor.2008.04.004}

\leavevmode\vadjust pre{\hypertarget{ref-randtoolbox2020}{}}%
Chalabi, Y., Dutang, C., Savicky, P., \& Wuertz, D. (2020). \emph{{randtoolbox: Toolbox for Pseudo and Quasi Random Number Generation and Random Generator Tests}}. \url{https://CRAN.R-project.org/package=randtoolbox}

\leavevmode\vadjust pre{\hypertarget{ref-cukier1978}{}}%
Cukier, R. I., Levine, H. B., \& Shuler, K. E. (1978). Nonlinear sensitivity analysis of multiparameter model systems. \emph{Journal of Computational Physics}, \emph{26}(1), 1--42. https://doi.org/\url{https://doi.org/10.1016/0021-9991(78)90097-9}

\leavevmode\vadjust pre{\hypertarget{ref-daveiga2021a}{}}%
Da Veiga, S. (2021). \emph{{Kernel-based ANOVA decomposition and Shapley effects -- Application to global sensitivity analysis}}. arXiv. \url{https://doi.org/10.48550/ARXIV.2101.05487}

\leavevmode\vadjust pre{\hypertarget{ref-daveiga2021}{}}%
Da Veiga, S., Gamboa, F., Iooss, B., \& Prieur, C. (2021). \emph{Basics and trends in sensitivity analysis: Theory and practice in r}. SIAM. \url{https://doi.org/10.1137/1.9781611976694}

\leavevmode\vadjust pre{\hypertarget{ref-DiceDesign}{}}%
Dupuy, D., Helbert, C., \& Franco, J. (2015). {DiceDesign and DiceEval: two R packages for design and analysis of computer experiments}. \emph{Journal of Statistical Software}, \emph{65}(11), 1--38. \url{https://doi.org/10.18637/jss.v065.i11}

\leavevmode\vadjust pre{\hypertarget{ref-efron1981}{}}%
Efron, B., \& Stein, C. (1981). The jackknife estimate of variance. \emph{The Annals of Statistics}, \emph{9}(3), 586--596. \url{https://doi.org/10.1214/aos/1176345462}

\leavevmode\vadjust pre{\hypertarget{ref-fort2021}{}}%
Fort, J.-C., Klein, T., \& Lagnoux, A. (2021). Global sensitivity analysis and {W}asserstein spaces. \emph{SIAM/ASA Journal on Uncertainty Quantification}, \emph{9}(2), 880--921. \url{https://doi.org/10.1137/20M1354957}

\leavevmode\vadjust pre{\hypertarget{ref-gan2014}{}}%
Gan, Y., Duan, Q., Gong, W., Tong, C., Sun, Y., Chu, W., Ye, A., Miao, C., \& Di, Z. (2014). A comprehensive evaluation of various sensitivity analysis methods: A case study with a hydrological model. \emph{Environmental Modelling \& Software}, \emph{51}, 269--285. https://doi.org/\url{https://doi.org/10.1016/j.envsoft.2013.09.031}

\leavevmode\vadjust pre{\hypertarget{ref-gilquin2019}{}}%
Gilquin, L., Arnaud, E., Prieur, C., \& Janon, A. (2019). Making the best use of permutations to compute sensitivity indices with replicated orthogonal arrays. \emph{Reliability Engineering \& System Safety}, \emph{187}, 28--39. https://doi.org/\url{https://doi.org/10.1016/j.ress.2018.09.010}

\leavevmode\vadjust pre{\hypertarget{ref-gilquin2017}{}}%
Gilquin, L., Jiménez Rugama, L. A., Arnaud, É., Hickernell, F. J., Monod, H., \& Prieur, C. (2017). {Iterative construction of replicated designs based on Sobol' sequences}. \emph{Comptes Rendus Mathematique}, \emph{355}(1), 10--14. https://doi.org/\url{https://doi.org/10.1016/j.crma.2016.11.013}

\leavevmode\vadjust pre{\hypertarget{ref-halton1960}{}}%
Halton, J. H. (1960). On the efficiency of certain quasi-random sequences of points in evaluating multi-dimensional integrals. \emph{Numerische Mathematik}, \emph{2}(1), 84--90. \url{https://doi.org/10.1007/BF01386213}

\leavevmode\vadjust pre{\hypertarget{ref-harenberg2019}{}}%
Harenberg, D., Marelli, S., Sudret, B., \& Winschel, V. (2019). Uncertainty quantification and global sensitivity analysis for economic models. \emph{Quantitative Economics}, \emph{10}(1), 1--41. https://doi.org/\url{https://doi.org/10.3982/QE866}

\leavevmode\vadjust pre{\hypertarget{ref-homma1996}{}}%
Homma, T., \& Saltelli, A. (1996). Importance measures in global sensitivity analysis of nonlinear models. \emph{Reliability Engineering \& System Safety}, \emph{52}(1), 1--17. https://doi.org/\url{https://doi.org/10.1016/0951-8320(96)00002-6}

\leavevmode\vadjust pre{\hypertarget{ref-iooss2019}{}}%
Iooss, B., \& Prieur, C. (2019). {Shapley effects for sensitivity analysis with correlated inputs: comparisons with Sobol' indices, numerical estimation and applications}. \emph{International Journal for Uncertainty Quantification}, \emph{9}(5), 493--514.

\leavevmode\vadjust pre{\hypertarget{ref-iooss2009}{}}%
Iooss, B., \& Ribatet, M. (2009). Global sensitivity analysis of computer models with functional inputs. \emph{Reliability Engineering \& System Safety}, \emph{94}(7), 1194--1204. https://doi.org/\url{https://doi.org/10.1016/j.ress.2008.09.010}

\leavevmode\vadjust pre{\hypertarget{ref-sensitiviti2021}{}}%
Iooss, B., Veiga, S. D., Janon, A., Gilles Pujol, with contributions from B. B., Boumhaout, K., Delage, T., Amri, R. E., Fruth, J., Gilquin, L., Guillaume, J., Idrissi, M., Gratiet, L. L., Lemaitre, P., Marrel, A., Meynaoui, A., Nelson, B. L., Monari, F., Oomen, R., Rakovec, O., \ldots{} Weber, F. (2021). \emph{{sensitivity: Global Sensitivity Analysis of Model Outputs}}. \url{https://CRAN.R-project.org/package=sensitivity}

\leavevmode\vadjust pre{\hypertarget{ref-ishigami1990}{}}%
Ishigami, T., \& Homma, T. (1990). An importance quantification technique in uncertainty analysis for computer models. \emph{{First International Symposium on Uncertainty Modeling and Analysis}}, 398--403. \url{https://doi.org/10.1109/ISUMA.1990.151285}

\leavevmode\vadjust pre{\hypertarget{ref-janon2014}{}}%
Janon, A., Klein, T., Lagnoux, A., Nodet, M., \& Prieur, C. (2014). Asymptotic normality and efficiency of two {Sobol} index estimators. \emph{ESAIM: Probability and Statistics}, \emph{18}, 342--364. \url{https://doi.org/10.1051/ps/2013040}

\leavevmode\vadjust pre{\hypertarget{ref-jansen1999}{}}%
Jansen, M. J. W. (1999). Analysis of variance designs for model output. \emph{Computer Physics Communications}, \emph{117}(1), 35--43. https://doi.org/\url{https://doi.org/10.1016/S0010-4655(98)00154-4}

\leavevmode\vadjust pre{\hypertarget{ref-kleijnen2009-1}{}}%
Kleijnen, J. P. C. (2009). \emph{Factor screening in simulation experiments: Review of sequential bifurcation} (C. Alexopoulos, D. Goldsman, \& J. R. Wilson, Eds.; Vol. 133, pp. 153--167). Springer. \url{https://doi.org/10.1007/b110059_8}

\leavevmode\vadjust pre{\hypertarget{ref-marrel2012}{}}%
Marrel, A., Iooss, B., Da Veiga, S., \& Ribatet, M. (2012). Global sensitivity analysis of stochastic computer models with joint metamodels. \emph{Statistics and Computing}, \emph{22}, 833--847. \url{https://doi.org/10.1007/s11222-011-9274-8}

\leavevmode\vadjust pre{\hypertarget{ref-mckay1979}{}}%
McKay, M. D., Beckman, R. J., \& Conover, W. J. (1979). A comparison of three methods for selecting values of input variables in the analysis of output from a computer code. \emph{Technometrics}, \emph{21}(2), 239--245.

\leavevmode\vadjust pre{\hypertarget{ref-mohammadi2019}{}}%
Mohammadi, H., Challenor, P., \& Goodfellow, M. (2019). {Emulating dynamic non-linear simulators using Gaussian processes}. \emph{Computational Statistics \& Data Analysis}, \emph{139}, 178--196. https://doi.org/\url{https://doi.org/10.1016/j.csda.2019.05.006}

\leavevmode\vadjust pre{\hypertarget{ref-niederreiter1992}{}}%
Niederreiter, H. (1992). \emph{{Random number generation and quasi-Monte Carlo methods}}. Society for Industrial; Applied Mathematics. \url{https://doi.org/10.1137/1.9781611970081}

\leavevmode\vadjust pre{\hypertarget{ref-ohagan2006}{}}%
O'Hagan, A. (2006). Bayesian analysis of computer code outputs: A tutorial. \emph{Reliability Engineering \& System Safety}, \emph{91}(10), 1290--1300. \url{https://doi.org/10.1016/j.ress.2005.11.025}

\leavevmode\vadjust pre{\hypertarget{ref-owen2014}{}}%
Owen, A. B. (2014). {Sobol' indices and Shapley value}. \emph{SIAM/ASA Journal on Uncertainty Quantification}, \emph{2}(1), 245--251. \url{https://doi.org/10.1137/130936233}

\leavevmode\vadjust pre{\hypertarget{ref-pianosi2016}{}}%
Pianosi, F., Beven, K., Freer, J., Hall, J. W., Rougier, J., Stephenson, D. B., \& Wagener, T. (2016). Sensitivity analysis of environmental models: A systematic review with practical workflow. \emph{Environmental Modelling \& Software}, \emph{79}, 214--232. https://doi.org/\url{https://doi.org/10.1016/j.envsoft.2016.02.008}

\leavevmode\vadjust pre{\hypertarget{ref-puy2021}{}}%
Puy, A., Piano, S. L., Saltelli, A., \& Levin, S. A. (2021). \emph{{sensobol: An R package to compute variance-based sensitivity indices}}. \url{https://arxiv.org/abs/2101.10103}

\leavevmode\vadjust pre{\hypertarget{ref-qian2020}{}}%
Qian, G., \& Mahdi, A. (2020). Sensitivity analysis methods in the biomedical sciences. \emph{Mathematical Biosciences}, \emph{323}, 108306. https://doi.org/\url{https://doi.org/10.1016/j.mbs.2020.108306}

\leavevmode\vadjust pre{\hypertarget{ref-qian2009}{}}%
Qian, P. Z. G. (2009). Nested latin hypercube designs. \emph{Biometrika}, \emph{96}(4), 957--970.

\leavevmode\vadjust pre{\hypertarget{ref-radaiideh2019}{}}%
Radaideh, M. I., Surani, S., O'Grady, D., \& Kozlowski, T. (2019). Shapley effect application for variance-based sensitivity analysis of the few-group cross-sections. \emph{Annals of Nuclear Energy}, \emph{129}, 264--279. https://doi.org/\url{https://doi.org/10.1016/j.anucene.2019.02.002}

\leavevmode\vadjust pre{\hypertarget{ref-GPML}{}}%
Rasmussen, C. E., \& Williams, C. K. I. (2005). \emph{Gaussian processes for machine learning (adaptive computation and machine learning)}. The MIT Press.

\leavevmode\vadjust pre{\hypertarget{ref-razavi2021}{}}%
Razavi, S., Jakeman, A., Saltelli, A., Prieur, C., Iooss, B., Borgonovo, E., Plischke, E., Lo Piano, S., Iwanaga, T., Becker, W., Tarantola, S., Guillaume, J. H. A., Jakeman, J., Gupta, H., Melillo, N., Rabitti, G., Chabridon, V., Duan, Q., Sun, X., \ldots{} Maier, H. R. (2021). The future of sensitivity analysis: An essential discipline for systems modeling and policy support. \emph{Environmental Modelling \& Software}, \emph{137}, 104954. https://doi.org/\url{https://doi.org/10.1016/j.envsoft.2020.104954}

\leavevmode\vadjust pre{\hypertarget{ref-roustant2012}{}}%
Roustant, O., Ginsbourger, D., \& Deville, Y. (2012). {DiceKriging}, {DiceOptim}: Two {R} packages for the analysis of computer experiments by kriging-based metamodeling and optimization. \emph{Journal of Statistical Software}, \emph{51}(1), 1--55. \url{https://doi.org/10.18637/jss.v051.i01}

\leavevmode\vadjust pre{\hypertarget{ref-saltelli2002}{}}%
Saltelli, Andrea. (2002). Making best use of model evaluations to compute sensitivity indices. \emph{Computer Physics Communications}, \emph{145}(2), 280--297. https://doi.org/\url{https://doi.org/10.1016/S0010-4655(02)00280-1}

\leavevmode\vadjust pre{\hypertarget{ref-saltelli2004}{}}%
Saltelli, Andrea, Tarantola, S., Campolongo, F., \& Ratto, M. (2004). \emph{{Sensitivity analysis in practice: A guide to assessing scientific models}}. Wiley. \url{https://doi.org/10.1002/0470870958}

\leavevmode\vadjust pre{\hypertarget{ref-saltelli1999}{}}%
Saltelli, A., Tarantola, S., \& Chan, K. P.-S. (1999). A quantitative model-independent method for global sensitivity analysis of model output. \emph{Technometrics}, \emph{41}(1), 39--56. \url{https://doi.org/10.1080/00401706.1999.10485594}

\leavevmode\vadjust pre{\hypertarget{ref-santner2003}{}}%
Santner, T. J., B., W., \& W., N. (2003). \emph{The design and analysis of computer experiments} (p. 283). Springer-Verlag.

\leavevmode\vadjust pre{\hypertarget{ref-shapley1953}{}}%
Shapley, L. S. (1953). A value for n-person games {(AM-28)}. In H. W. Kuhn \& A. W. Tucker (Eds.), \emph{{Contributions to the theory of games, Volume II}} (pp. 307--318). Princeton University Press.

\leavevmode\vadjust pre{\hypertarget{ref-sobol1967}{}}%
Sobol', I. M. (1967). On the distribution of points in a cube and the approximate evaluation of integrals. \emph{USSR Computational Mathematics and Mathematical Physics}, \emph{7}(4), 86--112. https://doi.org/\url{https://doi.org/10.1016/0041-5553(67)90144-9}

\leavevmode\vadjust pre{\hypertarget{ref-sobol2001}{}}%
Sobol', I. M. (2001). Global sensitivity indices for nonlinear mathematical models and their monte carlo estimates. \emph{Mathematics and Computers in Simulation}, \emph{55}(1), 271--280. https://doi.org/\url{https://doi.org/10.1016/S0378-4754(00)00270-6}

\leavevmode\vadjust pre{\hypertarget{ref-song2016}{}}%
Song, E., Nelson, B. L., \& Staum, J. (2016). Shapley effects for global sensitivity analysis: Theory and computation. \emph{SIAM/ASA Journal on Uncertainty Quantification}, \emph{4}(1), 1060--1083. \url{https://doi.org/10.1137/15M1048070}

\leavevmode\vadjust pre{\hypertarget{ref-stein1987}{}}%
Stein, M. (1987). Large sample properties of simulations using {L}atin hypercube sampling. \emph{Technometrics}, \emph{29}(2), 143--151.

\leavevmode\vadjust pre{\hypertarget{ref-sudret2008}{}}%
Sudret, B. (2008). Global sensitivity analysis using polynomial chaos expansions. \emph{Reliability Engineering \& System Safety}, \emph{93}(7), 964--979. https://doi.org/\url{https://doi.org/10.1016/j.ress.2007.04.002}

\leavevmode\vadjust pre{\hypertarget{ref-sun2011}{}}%
Sun, Y., Apley, D. W., \& Staum, J. (2011). Efficient nested simulation for estimating the variance of a conditional expectation. \emph{Operations Research}, \emph{59}(4), 998--1007. \url{https://doi.org/10.1287/opre.1110.0932}

\leavevmode\vadjust pre{\hypertarget{ref-tissot2015}{}}%
Tissot, J.-Y., \& Prieur, C. (2015). {A randomized orthogonal array-based procedure for the estimation of first- and second-order Sobol' indices}. \emph{Journal of Statistical Computation and Simulation}, \emph{85}(7), 1358--1381. \url{https://doi.org/10.1080/00949655.2014.971799}

\leavevmode\vadjust pre{\hypertarget{ref-vernon2018}{}}%
Vernon, I., Liu, J., Goldstein, M., Rowe, J., Topping, J., \& Lindsey, K. (2018). Bayesian uncertainty analysis for complex systems biology models: Emulation, global parameter searches and evaluation of gene functions. \emph{BMC Systems Biology}, \emph{12}(1), 1. \url{https://doi.org/10.1186/s12918-017-0484-3}

\leavevmode\vadjust pre{\hypertarget{ref-ggplot2}{}}%
Wickham, H. (2016). \emph{{ggplot2: Elegant Graphics for Data Analysis}}. Springer-Verlag New York. \url{https://ggplot2.tidyverse.org}

\leavevmode\vadjust pre{\hypertarget{ref-zhang2013}{}}%
Zhang, C., Chu, J., \& Fu, G. (2013). {Sobol's sensitivity analysis for a distributed hydrological model of Yichun River Basin, China}. \emph{Journal of Hydrology}, \emph{480}, 58--68. https://doi.org/\url{https://doi.org/10.1016/j.jhydrol.2012.12.005}

\leavevmode\vadjust pre{\hypertarget{ref-zhu2021}{}}%
Zhu, X., \& Sudret, B. (2021). Global sensitivity analysis for stochastic simulators based on generalized lambda surrogate models. \emph{Reliability Engineering \& System Safety}, \emph{214}, 107815. https://doi.org/\url{https://doi.org/10.1016/j.ress.2021.107815}

\end{CSLReferences}

\end{document}